\documentclass[a4paper,10pt]{article}
\pdfoutput=1
\usepackage{jcappub}
\usepackage[T1]{fontenc}
\usepackage{lmodern}
\usepackage{mathtools}
\usepackage{microtype}
\usepackage{booktabs}
\usepackage[capitalize]{cleveref}

\newcommand{\Mpl}{M_\mathrm{pl}}
\newcommand{\abs}[1]{{\left \vert #1 \right \vert}}
\newcommand{\overbar}[1]{\mkern 1.25mu\overline{\mkern-1.25mu#1\mkern-1.25mu}\mkern 1.25mu}
\newcommand{\ud}{\mathrm{d}}
\newcommand{\dd}[2]{\frac{\ud {#1} }{\ud {#2}}}

\newcommand{\pd}[2]{\frac{\partial {#1} }{\partial {#2}}}

\newcommand{\deltafoverf}{\Psi}
\newcommand{\hD}{h}
\newcommand{\fnu}{\Omega_\nu}
\newcommand{\Rnu}{R_\nu^\mathrm{SM}}

\title{Probing neutrino interactions and dark radiation with gravitational waves}

\author{Marilena Loverde}
\author{and Zachary J. Weiner}

\affiliation{
    Department of Physics, University of Washington,\\
    Seattle, WA 98195, U.S.A.}

\emailAdd{mloverde@uw.edu}
\emailAdd{zweiner@uw.edu}

\abstract{
    After their generation, cosmological backgrounds of gravitational waves propagate nearly freely
    but for the expansion of the Universe and the anisotropic stress of free-streaming particles.
    Primordial signals---both that from inflation and the infrared spectrum associated
    to subhorizon production mechanisms---would carry clean information about the cosmological
    history of these effects.
    We study the modulation of the standard damping of gravitational waves by free-streaming
    radiation due to the decoupling (or recoupling) of interactions.
    We focus on nonstandard neutrino interactions in effect after the decoupling of weak
    interactions as well as more general scenarios in the early Universe involving other light
    relics.
    We develop semianalytic results in fully free-streaming scenarios to provide intuition for
    numerical results that incorporate interaction rates with a variety of temperature dependencies.
    Finally, we compute the imprint of neutrino interactions on the $B$-mode polarization of the
    cosmic microwave background, and we comment on other means to infer the presence of such effects
    at higher frequencies.
}

\begin{document}
\maketitle
\flushbottom

\section{Introduction}\label{sec:introduction}

Observing a background of low-frequency gravitational waves remains a top science goal for
contemporary and upcoming cosmic microwave background (CMB) experiments~\cite{CMB-S4:2022ght,
Chang:2022tzj, CMB-S4:2016ple, CMB-S4:2020lpa, SPT-3G:2014dbx, Kamionkowski:2015yta, Staggs:2018gvf,
SimonsObservatory:2018koc, Caprini:2018mtu, BICEP:2021xfz, Komatsu:2022nvu}.
A detection of primordial gravitational waves, on top of the confirmed predictions that the initial
conditions of large-scale density fluctuations would be adiabatic, nearly Gaussian, and nearly scale
invariant~\cite{Planck:2018jri, Planck:2018vyg}, would all but confirm an early phase of quasi de
Sitter expansion as the origin of structure in the Universe~\cite{Starobinsky:1979ty, Guth:1980zm,
Sato:1980yn, Mukhanov:1981xt, Linde:1981mu, Albrecht:1982wi, Linde:1983gd, Achucarro:2022qrl}.
In the simplest models of inflation, the amplitude of the gravitational wave spectrum directly
measures the (otherwise unknown) energy scale of the fundamental physics at play.
The latest constraints limit the amplitude of such tensor fluctuations to be smaller than $0.036$
times its scalar counterpart at the 95\% confidence level~\cite{BICEP:2021xfz}, corresponding to an
upper energy scale of $1.4 \times 10^{16} \, \mathrm{GeV}$.
If the energy scale of inflation is not dramatically smaller than this current limit, then upcoming
experiments could observe the signature of primordial gravitational waves in the $B$-mode
(curl-like) polarization of the CMB anisotropies.

At the same time, interferometric experiments and pulsar timing arrays have opened a new window on
relativistic astrophysical phenomena---principally via gravitational wave signals from mergers of
compact objects or black holes~\cite{LIGOScientific:2016aoc,NANOGrav:2020bcs,Ballmer:2022uxx}.
For the cosmologist, aside from the potential to probe phase transitions~\cite{Kosowsky:1991ua,
Kosowsky:1992rz, Kamionkowski:1993fg, Grojean:2006bp, Caprini:2015zlo, Caprini:2019egz}, topological
defects~\cite{Kibble:1976sj, Vilenkin:1981zs, Vachaspati:1984gt, Damour:2004kw,
Blanco-Pillado:2017rnf, Auclair:2019wcv, LIGOScientific:2021nrg}, particle
production~\cite{Kitajima:2018zco, Machado:2018nqk, Arvanitaki:2019rax, Weiner:2020sxn,
Cui:2021are}, and primordial black holes~\cite{Nakamura:1997sm, Saito:2008jc, Sasaki:2016jop,
Bird:2016dcv, Sasaki:2018dmp, Carr:2020gox}, direct detection experiments offer a high-frequency
anchor which could provide complementary information about the possible primordial background from
inflation~\cite{Smith:2005mm, Lasky:2015lej, Seto:2001qf, Crowder:2005nr, Corbin:2005ny,
Harry:2006fi, Kawamura:2006up, Yagi:2011wg, Isoyama:2018rjb, Carilli:2004nx, Janssen:2014dka,
Weltman:2018zrl}.
By constraining gravitational waves over a large range in frequency (at scales far smaller than
those that would imprint upon the observable CMB), gravitational wave experiments could provide a
unique view into the physics of the very early Universe (see Refs.~\cite{Maggiore:1999vm,
Caprini:2018mtu, Caldwell:2022qsj} for reviews).

Measurements of inflationary gravitational waves would inform more physics than just that of
inflation itself.
The present-day spectrum is sensitive to the thermal history of the Standard Model plasma and the
expansion history of the Universe---especially to large departures from radiation
domination~\cite{Grishchuk:1974ny, Giovannini:1998bp, Boyle:2005se, Kuroyanagi:2011fy, Li:2013nal,
Saikawa:2018rcs, Figueroa:2019paj}.
Moreover, cosmological gravitational waves also interact with relativistic particles.
Once the weak interactions become inefficient (when the Universe was about a second old), Standard
Model neutrinos begin free streaming, reducing the amplitude of inflationary gravitational waves by
about $20\%$ on scales larger than the horizon at that time~\cite{Weinberg:2003ur}.
In general, gravitational waves probe the interaction history of the entire relativistic content of
the Universe.
In this paper we investigate the imprints of beyond-the-Standard-Model interactions of neutrinos (or
other hypothetical particles) on the stochastic gravitational wave background.

Within the Standard Model (SM), neutrinos are neutral and couple only via the weak interactions, and
many of their properties are largely inaccessible to current laboratory experiments.
Yet for much of the Universe's history neutrinos were the second-most abundant species, for which
reason cosmological observations provide substantial information about their dynamics.
After the weak interactions become inefficient, the cosmological role of neutrinos is purely
gravitational.
At least two of the neutrinos are massive, and cosmological data place upper limits on their
absolute mass scale via their contribution to the expansion rate and growth of structure when
nonrelativistic~\cite{deSalas:2020pgw, Esteban:2020cvm}.
At earlier times when they are relativistic, the observable signatures of spatial perturbations to
the neutrino distribution is predominantly governed by whether they scatter frequently (and so are
fluidlike) or interact rarely (and instead free stream)~\cite{Bashinsky:2003tk}.
In the latter case, their anisotropic stress not only impacts baryon acoustic
oscillations~\cite{Bashinsky:2003tk,Hou:2011ec,Baumann:2015rya,Choi:2018gho} but is also responsible
for the aforementioned damping of gravitational waves.
Measurements of the temperature anisotropies of the CMB confirm that most (if not all) of the energy
density in neutrinos is
free-streaming~\cite{Baumann:2015rya,Follin:2015hya,Blinov:2020hmc,Brinckmann:2020bcn};
however, it remains possible that neutrinos have self-interactions that decouple at some time around
recombination~\cite{Cyr-Racine:2013jua,Lancaster:2017ksf,Oldengott:2017fhy,Barenboim:2019tux,Brinckmann:2020bcn,Das:2020xke,Kreisch:2022zxp,Taule:2022jrz,RoyChoudhury:2022rva}.

Relativistic particles interact not only with the superhorizon gravitational waves frozen out during
inflation but also with the low-frequency tail of stochastic backgrounds generated by subhorizon
processes~\cite{Hook:2020phx, Brzeminski:2022haa}.
A promising candidate is that from the electroweak phase transition, which could be detected by
LISA if the transition is first order~\cite{Caprini:2019egz}.
Free-streaming, relativistic particles would leave a characteristic signature on the infrared,
``causal'' part of the spectrum, altering its power-law dependence or even inducing oscillatory
features.
The dynamics of causal gravitational waves are largely agnostic to the details of the underlying
source, effectively reducing to a transfer function encoding propagation effects to the present day,
similar to that applicable to inflationary gravitational waves.
We also investigate the imprints of self-interacting radiation in this complementary scenario.

In the remainder of this paper, we first present the theory of cosmological perturbations in
application to relativistic degrees of freedom and gravitational waves (in \cref{sec:background}).
In \cref{sec:inflation} we develop semianalytic solutions for the effect of noninteracting
neutrinos on primordial gravitational waves from inflation, building intuition with which to
understand numerical results for interacting neutrinos.
We then consider ``causal'' gravitational wave backgrounds from subhorizon sources in
\cref{sec:causal}.
These two scenarios are mathematically distinguished by the gravitational waves' initial
conditions---a frozen-out amplitude or a sudden jump in velocity, respectively---each exhibiting a
starkly different response to free-streaming radiation.
Because the anisotropic stress sourced by collisionless radiation decays with expansion inside the
horizon, the causal and inflationary scenarios together provide a nearly complete treatment of the
effect of free-streaming radiation on gravitational waves.
Though we focus on the effect of neutrinos for concreteness, our analysis extends straightforwardly
to other, hypothetical dark radiation species.\footnote{
    The rest mass of neutrinos or other dark radiation is irrelevant so long the particles are
    relativistic while the scales of interest enter the horizon.
}
In \cref{sec:discussion} we conclude with a discussion of future prospects for constraining neutrino
interactions and hypothetical new degrees of freedom in CMB $B$-modes and direct gravitational wave
measurements.
\Cref{app:numerical-implementation} details our numerical implementation, and
\cref{app:semianalytic-inflation} outlines a semianalytic calculation of the damping effect
employed in \cref{sec:inflation}.

\section{Gravitational waves and relativistic particles}\label{sec:background}

We first review the well-established theory of the dynamics of gravitational and matter
perturbations in the early Universe in application to the coupling of gravitational waves and
relativistic species (of any spin).
Gravitational waves also interact with other types of matter, including relativistic
axions~\cite{Dent:2013asa, Ringwald:2020vei}, nonrelativistic and collisional
matter~\cite{Baym:2017xvh, Flauger:2017ged, Miron-Granese:2020hyq, Zarei:2021dpb}, and vector
fields~\cite{Bielefeld:2015daa, Miravet:2020kuj, Tishue:2021blv, Miravet:2022pli}; some of these
scenarios are not captured by the kinetic theory treatment we review.
We neglect any chemical potential that would be requisite to describe, e.g., chiral fermions and
polarized gravitational waves~\cite{Ichiki:2006rn, Valle:2013aia, Barrie:2017mmr, Sadofyev:2017zqc,
Gubler:2022zmf}.

\subsection{Cosmological perturbation theory}

Since we are interested only in gravitational wave observables, we consider a
Friedmann-Lema\^itre-Robertson-Walker (FLRW) spacetime perturbed only by a tensor mode $\hD_{ij}$:
\begin{align}\label{eqn:background-flrw-metric}
    \ud s^2
    &= a(\tau)^2 \left(
            - \ud \tau^2
            + \left[ \delta_{ij} + \hD_{ij}(\tau, \mathbf{x}) \right] \ud x^i \ud x^j
        \right).
\end{align}
Primes indicate derivatives with respect to the conformal time $\tau$, and we denote the conformal
Hubble parameter as $\mathcal{H}(\tau) \equiv a'(\tau) / a(\tau)$, in terms of which the (standard)
Hubble parameter is $H(\tau) = \mathcal{H}(\tau) / a(\tau)$.
Repeated spatial (Latin) indices are contracted with the Kronecker delta function regardless of
their placement.
The tensor perturbation $\hD_{ij}$ is transverse ($\partial_i \hD_{ij} = 0$) and traceless
($\hD_{ii} = 0$); its Fourier modes may therefore be expanded in terms of two polarizations
$\lambda$ as
\begin{align}
    \hD_{ij}(\tau, \mathbf{x})
    = \int \frac{\ud^3 k}{(2 \pi)^3} \hD_{ij}(\tau, \mathbf{k}) e^{i \mathbf{k} \cdot \mathbf{x}}
    &\equiv \int \frac{\ud^3 k}{(2 \pi)^3} e^{i \mathbf{k} \cdot \mathbf{x}}
        \sum_\lambda \hD_\lambda(\tau, \mathbf{k}) \epsilon^\lambda_{ij}(\mathbf{k}).
    \label{eqn:def-hij-pol-expansion}
\end{align}
We do not have cause to pick a particular basis of polarization tensors
$\epsilon^\lambda_{ij}(\mathbf{k})$, but any choice must likewise be transverse and traceless.
The Einstein equation for $\hD_{ij}$, decomposed onto this basis, takes the form of two
inhomogeneous, damped waved equations,
\begin{align}\label{eqn:h-lambda-eom}
    \hD_\lambda'' + 2 \mathcal{H} \hD_\lambda' - \partial_k \partial_k \hD_\lambda
    &= \frac{2 a^2}{\Mpl^2} \pi^T_\lambda.
\end{align}
Gravitational waves are sourced by the transverse and traceless component $\pi^T_{ij}$ of the
perturbation to the stress tensor, $\delta T^{i}_{\hphantom{i}j}$.

Gravitational waves carry an effective energy density which, deep inside the horizon (i.e.,
$k \gg \mathcal{H}$), is~\cite{Abramo:1997hu,Brandenberger:2018fte,Clarke:2020bil}
\begin{align}
    \bar{\rho}_\mathrm{GW}(\tau)
    = \frac{\Mpl^2}{8 a^2} \left\langle
            \hD_{ij}' \hD_{ij}'
            + \partial_k \hD_{ij} \partial_k \hD_{ij}
        \right\rangle,
    \label{eqn:rho-gw-subhorizon}
\end{align}
where the angled brackets a spatial average.
The observable of interest is the fractional contribution by gravitational waves of a given
wavenumber to the background energy density of the Universe,
$\bar{\rho}(\tau) = 3 H(\tau)^2 \Mpl^2$.
We therefore define the spectral abundance of gravitational waves,
\begin{align}\label{eqn:omega-gw-spectrum-def}
    \Omega_\mathrm{GW}(\tau, k)
	= \frac{1}{\bar{\rho}(\tau)} \dd{\bar{\rho}_\mathrm{GW}(\tau)}{\ln k}.
\end{align}
Since gravitational waves deep inside the horizon are effectively undamped harmonic oscillators,
$\abs{\hD_\lambda'(\tau, \mathbf{k})} \approx \abs{k \hD_\lambda(\tau, \mathbf{k})}$.
The effective energy density \cref{eqn:rho-gw-subhorizon} may then be expressed as an integral
over wavenumber by substituting the inverse Fourier transform of $\hD_{ij}$,
\cref{eqn:def-hij-pol-expansion}.
In terms of the dimensionless power spectrum of $\hD_\lambda$,
\begin{align}
    \langle
        \hD_{\lambda_1}(\tau, \mathbf{k}_1)
        \hD_{\lambda_2}(\tau, \mathbf{k}_2)
    \rangle
    &\equiv (2 \pi)^3 \delta^3(\mathbf{k}_1 + \mathbf{k}_2)
        \delta_{\lambda_1 \lambda_2}
        \frac{2 \pi^2}{k_1^3}
        \Delta^2_{\lambda_1}(\tau, k_1),
\end{align}
we may therefore write the spectral abundance of gravitational waves as
\begin{align}\label{eqn:omega-gw-spectrum-ito-power-spectrum}
    \Omega_\mathrm{GW}(\tau, k)
	= \frac{1}{12} \left( \frac{k}{\mathcal{H}(\tau)} \right)^2
        \sum_\lambda \overbar{\Delta^2_\lambda(\tau, k)}.
\end{align}
where the overbar denotes a time average (i.e., over oscillations to ensure the validity of
taking $\abs{\hD_\lambda'} \approx \abs{k \hD_\lambda}$).
The spectrum evaluated in the early Universe is related to that at the present day (at $\tau_0$) by
the transfer function~\cite{Caprini:2018mtu,Saikawa:2018rcs,Kite:2021yoe}
\begin{align}
    \Omega_{\mathrm{GW}}(\tau_0, k) h^2
    &= \Omega_{\mathrm{rad}}(\tau_0) h^2
        \frac{g_{\star}(\tau)}{g_{\star}(\tau_0)}
        \left( \frac{g_{\star S}(\tau)}{g_{\star S}(\tau_0)} \right)^{-4/3}
        \Omega_{\mathrm{GW}}(\tau, k)
    \label{eqn:gw-amplitude-transfer-function}
\end{align}
and would be observed at present-day frequencies related to wavenumber $k$ by
\begin{align}
    f
    = \frac{k / 2 \pi a(\tau)}{\sqrt{H(\tau) \Mpl}}
        \left[
            \Omega_{\mathrm{rad}}(\tau_0)
            H_0^2 \Mpl^2
        \right]^{1/4}
        \left( \frac{g_{\star}(\tau)}{g_{\star}(\tau_0)} \right)^{1/4}
        \left( \frac{g_{\star S}(\tau)}{g_{\star S}(\tau_0)} \right)^{-1/3}.
\end{align}
Here $g_{\star}$ and $g_{\star S}$ are the numbers of relativistic degrees of freedom in energy and
entropy density, respectively.
Note that the present-day abundance of radiation, evaluated as if neutrinos are massless, is
$\Omega_{\mathrm{rad}}(\tau_0) h^2 \approx 4.2 \times 10^{-5}$~\cite{Planck:2018vyg} and that
$H_0 / h \equiv 100 \, \mathrm{km} \, \mathrm{s}^{-1} / \mathrm{Mpc} \approx 3.24 \times 10^{-18} \, \mathrm{Hz}$.

Making further progress requires specifying the matter content that contributes to the anisotropic
stress tensor $\pi^T_\lambda$.
Following the standard prescription~\cite{Ma:1995ey}, we solve the Boltzmann equation for the phase
space density of particles, expanded into background and perturbed components as
\begin{align}
    f(\tau, \mathbf{x}, q, \hat{q})
    &\equiv \bar{f}(\tau, q)
        \left[ 1 + \deltafoverf(\tau, \mathbf{x}, q, \hat{q}) \right].
\end{align}
Like Ref.~\cite{Ma:1995ey}, we parameterize the functional dependence of the distribution function
in terms of spacetime $\mathbf{x}$ and $\tau$, comoving momentum $q = a p$ (where $p$ is the
magnitude of the proper momentum), and propagation direction $\hat{q}_i \equiv q_i / q$.
We also define the comoving energy $\mathcal{E} = a E$ in terms of the particle energy
$E = \sqrt{p^2 + m^2}$.
While we consider only relativistic particles with $p \gg m$, for which $\mathcal{E} = q$, we retain
the distinction for the time being.
The Boltzmann equation is
\begin{align}
    \dd{}{\tau} f(\tau, \mathbf{x}, q, \hat{q})
    &= \mathcal{C}[f],
\end{align}
where $\mathcal{C}[f]$ denotes the collision term.
In terms of $\deltafoverf$, the perturbation equation reads
\begin{align}
      \dd{}{\tau} \deltafoverf(\tau, \mathbf{x}, q, \hat{q})
        + \pd{\ln \bar{f}(\tau, q)}{q} \left( \dd{q}{\tau} \right)^{(1)}
        + \deltafoverf(\tau, \mathbf{x}, q, \hat{q}) \dd{\ln \bar{f}(\tau, q)}{\tau}
    &= \frac{1}{\bar{f}(\tau, q)} \mathcal{C}[f]^{(1)},
\end{align}
where a superscript $(1)$ denotes the perturbation to a given term.
In most relevant scenarios we may drop the term proportional to $\ud \bar{f} / \ud \tau$,
justified by noting that $\ud q / \ud \tau$ vanishes at the background level and assuming that the
background distribution function is a function of $q$ alone.
The latter condition holds, e.g., for relativistic species in equilibrium and for collisionless
relativistic or nonrelativistic particles (like SM neutrinos after weak decoupling).

Generically, $\deltafoverf$ comprises scalar, vector, and tensor contributions, but only the latter
sources gravitational waves.
Expanding the $\hat{q}$-dependence of the distribution function in Fourier space as
\begin{align}\label{eqn:deltafoverf-svt-decomposition}
    \deltafoverf(\tau, \mathbf{k}, q, \hat{q})
    &= \deltafoverf^{(S)}(\tau, \mathbf{k}, q, \hat{q})
        + \sum_{\lambda} \hat{q}_i \epsilon_i^\lambda(\mathbf{k})
            \deltafoverf^{(V)}_\lambda(\tau, \mathbf{k}, q, \hat{q})
        + \sum_{\lambda} \hat{q}_i \hat{q}_j \epsilon_{ij}^\lambda(\mathbf{k})
            \deltafoverf^{(T)}_\lambda(\tau, \mathbf{k}, q, \hat{q})
\end{align}
[with $\epsilon_i^\lambda(\mathbf{k})$ a suitable basis of transverse polarization vectors] defines
the scalar-vector-tensor decomposition of $\deltafoverf$ that uniquely maps into those components
that respectively source the scalar, vector, and tensor parts of Einstein's equations.\footnote{
    Seeing this requires observing in the geodesic equation that contributions from scalar, vector,
    and tensor perturbations to metric are each proportional to matching factors of
    $\hat{q}_i$, $\epsilon_i^\lambda$, and $\epsilon_{ij}^\lambda$.
    The linear-order Boltzmann equation may then be split into three independent equations on this
    basis.
    The scalar-vector-tensor decomposition of the stress-energy tensor [defined below in
    \cref{eqn:stress-tensor-def}] involves integrals over particle propagation directions
    weighted by various factors of $\hat{q}_i$ which reveal that
    \cref{eqn:deltafoverf-svt-decomposition} also provides the relevant decomposition for the
    Einstein equations.
}
To linear order in perturbations, the part of the Boltzmann equation proportional to
$\hat{q}_i \hat{q}_j \epsilon_{ij}^\lambda(\mathbf{k})$ is
\begin{align}
    \pd{\deltafoverf^{(T)}_\lambda}{\tau}
        + i k \mu \frac{q}{\mathcal{E}} \deltafoverf^{(T)}_\lambda
        - \frac{1}{2} \pd{\ln \bar{f}}{\ln q} \hD_\lambda'
    &= \frac{1}{\bar{f}(q)} \mathcal{C}[f]^{(T)}_\lambda.
    \label{eqn:ddeltafoverf-dtau-tensor}
\end{align}
Here $\mu = \hat{k} \cdot \hat{q}$ is the cosine of the angle between the wavenumber $\mathbf{k}$
and the propagation direction $\mathbf{q}$, and $\mathcal{C}[f]^{(T)}_\lambda$ denotes the tensor
component of the (yet-to-be-specified) collision term.
Because $\ud q / \ud \tau$ and $\ud \hat{q} / \ud \tau$ are themselves first order in perturbations
(via the geodesic equation), the terms proportional to
$\partial \deltafoverf^{(T)}_\lambda / \partial q$ and
$\partial \deltafoverf^{(T)}_\lambda / \partial \hat{q}_i$ vanish at leading order.

The angular ($\mu)$ dependence of the Boltzmann equation is most conveniently decomposed in a
partial wave expansion, i.e., onto the orthogonal basis of Legendre polynomials, $P_l(\mu)$.
This procedure recasts the Boltzmann equation into an infinite hierarchy of coupled equations for
the moments of the distribution function.
While one may substitute such an expansion for $\deltafoverf^{(T)}_\lambda$ directly, our focus on
relativistic species enables the $q$-dependence of the equations to be integrated out of the
Liouville operator [i.e., the left-hand side of \cref{eqn:ddeltafoverf-dtau-tensor}].
We therefore define the moment expansion of the (suitably normalized) perturbations to the
distribution function and the collision term as
\begin{align}\label{eqn:def-capital-F-ito-deltafoverf}
    F(\tau, \mathbf{k}, \hat{q})
    &\equiv \frac{
            \int \ud q \, q^3 \bar{f}(q) \deltafoverf(\tau, \mathbf{k}, q, \hat{q})
        }{
            \int \ud q \, q^3 \bar{f}(q)
        }
    \equiv \sum_{l = 0}^\infty (- i)^l \left( 2 l + 1 \right)
        F_l(\tau, \mathbf{k}) P_l(\mu) \\
\intertext{and}
    C(\tau, \mathbf{k}, \hat{q})
    &\equiv \frac{
            \int \ud q \, q^3 \mathcal{C}[f]^{(1)} / \bar{f}(q)
        }{
            \int \ud q \, q^3 \bar{f}(q)
        }
    \equiv \sum_{l = 0}^\infty (- i)^l \left( 2 l + 1 \right)
        C_l(\tau, \mathbf{k}) P_l(\mu).
    \label{eqn:def-collision-perturbation-and-hierarchy}
\end{align}
Inserting this expansion into \cref{eqn:ddeltafoverf-dtau-tensor} yields the Boltzmann hierarchy,
\begin{align}\label{eqn:collisions-boltzmann-F-T}
    \pd{F^{(T)}_{\lambda, l}}{\tau}
        - \frac{k}{2 l + 1}
        \left[
            l F^{(T)}_{\lambda, {l-1}}
            - \left( l + 1 \right) F^{(T)}_{\lambda, {l+1}}
        \right]
        + 2 \delta_{l0} \hD_\lambda'
    &= C_{\lambda, l}^{(T)},
\end{align}
after also taking the relativistic limit, $\mathcal{E} \to q$.
Though formally correct, \cref{eqn:collisions-boltzmann-F-T} is only useful if the collision term
has simple functional dependence on $F_{\lambda, l}$; this is the case for our treatment of
interactions (as discussed in \cref{sec:interactions}).

Knowing the evolution of the phase-space density itself, we now compute the source to (the tensor
part of) Einstein's equation.
The stress-energy tensor of a species with phase-space density $f$ is
\begin{align}\label{eqn:stress-tensor-def}
    T^{\alpha}_{\hphantom{\alpha}\beta}(\tau, \mathbf{k})
    &= \frac{1}{\sqrt{-g}}
        \int \ud p_1 \ud p_2 \ud p_3 \,
        \frac{p^\alpha p_\beta}{p^0}
        f(\tau, \mathbf{k}, q, \hat{q}),
\end{align}
with $p^\alpha$ the conjugate momentum.
For a relativistic species, the transverse-traceless component of the space-space perturbation
$\delta T^{i}_{\hphantom{i}j}$, projected onto $\epsilon^{\lambda}_{ij}(\mathbf{k})$,
is~\cite{Weinberg:2003ur,Weinberg:2008zzc}
\begin{align}\label{eqn:anisotropic-stress-ito-moments}
    \pi_\lambda^T(\tau, \mathbf{k})
    &= \bar{\rho}(\tau)
        \left(
            \frac{2}{15} F^{(T)}_{\lambda, 0}(\tau, \mathbf{k})
            + \frac{4}{21} F^{(T)}_{\lambda, 2}(\tau, \mathbf{k})
            + \frac{2}{35} F^{(T)}_{\lambda, 4}(\tau, \mathbf{k})
        \right).
\end{align}

The system of equations---\cref{eqn:h-lambda-eom} for $\hD_\lambda$ and
\cref{eqn:collisions-boltzmann-F-T} for $F^{(T)}_{\lambda, l}$---depend explicitly upon wavenumber
only via the combination $k \tau \equiv x$.
In terms of this dimensionless time coordinate, in Fourier space
\begin{subequations}\label{eqn:eoms-ito-x}
\begin{align}
    \partial_x F^{(T)}_{\lambda, l}
    &= \frac{1}{2 l + 1}
        \left[
            l F^{(T)}_{\lambda, {l-1}}
            - \left( l + 1 \right) F^{(T)}_{\lambda, {l+1}}
        \right]
        - 2 \delta_{l0} \partial_x \hD_\lambda
        + \frac{1}{k} C_{\lambda, l}^{(T)}
        \\
\intertext{and}
    \partial_x^2 \hD_\lambda
    &= - 2 \frac{\partial_x a}{a} \partial_x \hD_\lambda
        - \hD_\lambda
        + 6 \fnu(\tau) \left( \frac{\partial_x a}{a} \right)^2
        \left(
            \frac{2}{15} F^{(T)}_{\lambda, 0}
            + \frac{4}{21} F^{(T)}_{\lambda, 2}
            + \frac{2}{35} F^{(T)}_{\lambda, 4}
        \right).
    \label{eqn:h-eom-ito-x}
\end{align}
\end{subequations}
The fraction of energy in the relativistic species under consideration is
$\fnu(\tau) \equiv \bar{\rho}_\nu(\tau) / \bar{\rho}(\tau)$, suggestively labeled by $\nu$
(though we use this definition for any relativistic species under consideration, not just Standard
Model neutrinos).
Note that $\fnu(\tau)$ does not necessarily denote the fraction of energy in particles that are
specifically collisionless or free streaming at $\tau$---in interacting scenarios, the species $\nu$
(or some fraction thereof) may be fluidlike when its interactions are efficient.
In fully noninteracting scenarios, we refer to the fraction of energy in species that are
specifically free-streaming with $f_\mathrm{fs}$.

All that remains is to specify the background evolution of the Universe.
In a Universe dominated by a single component with equation of state $w$, the solution to the
Friedmann equations is
\begin{subequations}\label{eqn:single-component-solution}
\begin{align}
    \frac{a(\tau)}{a(\tau_i)}
    &= \left( \frac{\tau}{\tau_i} \right)^\alpha \\
    \mathcal{H}(\tau)
    &= \frac{\alpha}{\tau},
\end{align}
\end{subequations}
where $\alpha = 2 / (1 + 3 w)$.
In this case, $\partial_x a / a = \alpha / x$.
In a radiation Universe with $\alpha = 1$, $\fnu(\tau)$ is constant, barring decay channels into or
of the species $\nu$.
No species other than the photon and neutrinos are relativistic after electron-positron annihilation;
the neutrinos' share of the radiation energy is then
\begin{align}\label{eqn:R-nu-def}
    \Rnu
    &\equiv \frac{\bar{\rho}_\nu}{\bar{\rho}_\gamma + \bar{\rho}_\nu}.
\end{align}
For $N_\nu$ neutrino species,
\begin{align}\label{eqn:R-nu-standard-model}
    \Rnu
    &= \left[ 1 + \left( \frac{11}{4} \right)^{4/3} \frac{8}{7 N_\nu} \right]^{-1},
\end{align}
taking the value $\Rnu \approx 0.40523$ for $N_\nu = 3$.
Since the weak interactions decouple just as electron-positron annihilation begins, $\Rnu$
represents the free-streaming fraction during the radiation era (i.e., at temperatures
$T \lesssim 1 \, \mathrm{MeV}$ but well before matter-radiation equality).
When considering more general early-Universe scenarios, the free-streaming abundance $f_\mathrm{fs}$
may be treated as a free parameter.

More generally, we may employ the analytic solution for a matter-radiation Universe,
\begin{align}\label{eqn:scale-factor-solution-mr-universe}
    \frac{a(\tau)}{a(\tau_\mathrm{eq})}
    &= \frac{y}{8} \left( y + 4 \sqrt{2} \right)
\end{align}
where $y = k_\mathrm{eq} \tau$ and $k_\mathrm{eq} \equiv \mathcal{H}(\tau_\mathrm{eq}) \approx
10^{-2} \, \mathrm{Mpc}^{-1}$~\cite{Planck:2018vyg} is the horizon scale evaluated at
matter-radiation equality.
This scale corresponds to a present-day frequency
$f_\mathrm{eq} = 1.55 \times 10^{-17} \, \mathrm{Hz}$.
Note that $k_\mathrm{eq} \tau_\mathrm{eq} = 4 - 2 \sqrt{2} \approx 1.17$ and
\begin{align}\label{eqn:partial-y-a-over-a}
    \frac{\mathcal{H}(\tau)}{k_\mathrm{eq}}
    &= \frac{\partial_y a}{a}
    = \frac{4 \sqrt{2} + 2 y}{4 \sqrt{2} y + y^2}.
\end{align}
In this case, \cref{eqn:eoms-ito-x} depends on the ratio $k / k_\mathrm{eq}$, both via the change in
the expansion rate and because the neutrino abundance begins to decrease as the Universe enters
matter domination.
At all times after electron-positron annihilation,
\begin{align}\label{eqn:f-nu-vs-tau}
    \fnu(\tau)
    = \frac{\Rnu}{1 + a(\tau) / a(\tau_\mathrm{eq})}.
\end{align}
This background solution (neglecting dark energy) is sufficiently accurate to study the effect of
free-streaming radiation on all observable scales: the Universe expands by a factor more than $10^3$
by dark-energy--matter equality, with a correspondingly large reduction in $\fnu(\tau)$ [via
\cref{eqn:f-nu-vs-tau}] and thereby also the effect of free-streaming neutrinos on gravitational
waves.

\subsection{Modeling interacting radiation}\label{sec:interactions}

The elusiveness of neutrinos positions them as a prime candidate portal to dark sectors and physics
beyond the Standard Model.
Abundant motivations for novel interactions amongst the SM neutrinos include mechanisms for their
masses, extensions to the SM gauge sector, and the possibility of sterile neutrino dark matter (see
Refs.~\cite{Abazajian:2022ofy, Berryman:2022hds} for recent reviews).
Extensive literature has considered the cosmological signatures of neutrino
self-interactions~\cite{Raffelt:1987ah, Berkov:1987pz, Berkov:1988sd, Belotsky:2001fb,
Chacko:2003dt, Hannestad:2004qu, Hannestad:2005ex, Bell:2005dr, Cirelli:2006kt, Friedland:2007vv,
Basboll:2008fx, Bialynicka-Birula:1964ddi, Cyr-Racine:2013jua, Archidiacono:2013dua,
Archidiacono:2014nda, Forastieri:2015paa, Chu:2015ipa, Farzan:2015pca, Forastieri:2017oma,
Lancaster:2017ksf, Oldengott:2017fhy, Koksbang:2017rux, DiValentino:2017oaw, Song:2018zyl,
Kreisch:2019yzn, Barenboim:2019tux, Forastieri:2019cuf, Das:2020xke, RoyChoudhury:2020dmd,
Brinckmann:2020bcn, Esteban:2021ozz, Du:2021idh, Venzor:2022hql}, neutrino interactions with dark
matter~\cite{Mangano:2006mp, Serra:2009uu, Diacoumis:2017hff, Ghosh:2017jdy, Mosbech:2020ahp,
Paul:2021ewd, Green:2021gdc}, and interacting dark sectors more generally~\cite{Cyr-Racine:2012tfp,
Jeong:2013eza, Cyr-Racine:2013fsa, Buen-Abad:2015ova, Chacko:2015noa, Lesgourgues:2015wza,
Baumann:2015rya, Tang:2016mot, Ko:2016uft, Chacko:2016kgg, Prilepina:2016rlq, Brust:2017nmv,
Krall:2017xcw, Buen-Abad:2017gxg, Pan:2018zha, Chacko:2018vss, Choi:2018gho, Garny:2018byk,
Archidiacono:2019wdp, Bansal:2021dfh, Corona:2021qxl}.
A substantial amount of interest in these models is due to their potential to alleviate tensions in
measurements of the Hubble constant~\cite{DiValentino:2021izs} and the amplitude of matter
fluctuations---e.g., Refs.~\cite{Ko:2016uft, Chacko:2016kgg, Kumar:2017dnp, Buen-Abad:2017gxg,
DiValentino:2017oaw, Kreisch:2019yzn, Blinov:2019gcj, Archidiacono:2019wdp, Ghosh:2019tab,
Escudero:2019gvw, Blinov:2020hmc, He:2020zns, Berbig:2020wve, Becker:2020hzj, Choi:2020pyy,
Das:2020xke, Mazumdar:2020ibx, RoyChoudhury:2020dmd, Brinckmann:2020bcn, Bansal:2021dfh,
Aloni:2021eaq, Joseph:2022jsf, Buen-Abad:2022kgf}.

We now discuss the implementation of neutrino interactions (or that of other relativistic species),
i.e., the form of the collision term in the Boltzmann equation.
Under the assumption that the phase space perturbations are independent of momentum, the full
(momentum-dependent) collision term for $2-2$ scattering integrates to~\cite{Oldengott:2017fhy}
\begin{align}\label{eqn:collision-term-opacity}
    C_{\lambda, l}
    &= \alpha_l \partial_\tau \kappa_\nu F_{\lambda, l}.
\end{align}
Here $\kappa_\nu$ is the optical depth of the species $\nu$ (and $\partial_\tau \kappa_\nu$ the
time-dependent interaction rate), and the $\alpha_l$ are numerical coefficients that arise when
computing the moments of the full collision term.
This treatment is also referred to as the relaxation time approximation.

In principle, spatial fluctuations in phase space could be nonuniform over $q$.
Ref.~\cite{Oldengott:2017fhy} implemented the exact (momentum-dependent) scalar Boltzmann hierarchy
and the momentum-integrated hierarchy using \cref{eqn:collision-term-opacity} for the case of $2-2$
neutrino scattering mediated by a heavy scalar, finding that the dynamics of the neutrino fluid
variables and the resulting CMB angular power spectra agree extremely well between the two approaches.
A possible explanation for this result is that gravitational consequences of neutrinos (which are
the only cosmologically relevant ones after the weak interactions decouple) are themselves
momentum-integrated quantities and therefore are not sensitive to neutrino spectral
distortions~\cite{Cyr-Racine:2013jua}.
Furthermore, if the distribution function evolves substantially at the background level, then the
coefficients $\alpha_l$ would be time-dependent.
For instance, neutrinos could decay into a light scalar mediating self-interactions (that
``recouple'' below some energy scale); however, the produced mediator particles would also strongly
interact and be fluidlike.
Our description would then apply to the joint neutrino-mediator sector.
The $\alpha_l$ could still vary with time if the shape of the distribution function (over which the
collision term is integrated) evolves.
For simplicity to explore phenomenology, we neglect such model-dependent effects.

While the validity of \cref{eqn:collision-term-opacity} has not been quantitatively validated for,
e.g., neutrino interactions via a light mediator, nor explicitly for the tensor hierarchy under
study here, we assume it provides a sufficient approximation to capture the transition from
fluidlike to free-streaming behavior.
Ref.~\cite{Oldengott:2017fhy} computed the coefficients $\alpha_l$ for the scalar
Boltzmann hierarchy directly from the exact collision integral.
Energy and momentum conservation require that $\alpha_0$ and $\alpha_1$ vanish, while the values for
larger multipoles rise from $\alpha_2 = 0.40$ to $\alpha_l = 0.48$ for $l \geq 6$.
Noting that there is no conservation equation for tensors (i.e., the linear-order part of the
energy-momentum conservation equation, $\nabla_\mu T^{\mu \nu} = 0$, has no tensor component), we
take $\alpha_l = 1$ for all $l$; their exact values would only amount to an order unity change in
the relationship between the neutrino interaction rate and the time of de-/recoupling.

For generality's sake, we parameterize the interaction rate with temperature dependence of the form
\begin{align}\label{eqn:interaction-rate-general}
    \partial_\tau \kappa_\nu(\tau)
    &= - \frac{a(\tau)}{a(\tau_\star)} \lambda T_\nu(\tau)^n
\end{align}
in terms of an effective coupling constant $\lambda$ (with mass dimension $1 - n$), with
$a(\tau_\star)$ the scale factor at the time of the decoupling/recoupling transition.
Interactions are efficient when their rate is larger than the expansion rate, so
$\abs{\partial_\tau \kappa_\nu(\tau_\star)} = \mathcal{H}(\tau_\star)$ marks the time of the
transition.
Since both the comoving Hubble rate and the temperature decay as $1/a(\tau)$ in the radiation era,
decoupling occurs for scenarios with $n > 2$ and recoupling when $n < 2$.
Self-interactions mediated by heavy and light degrees of freedom, for example, respectively have
$n = 5$ and $n = 1$.
See Ref.~\cite{Taule:2022jrz} for a similar phenomenological approach applied to the effect of
neutrinos interactions on the CMB and large-scale structure.\footnote{
    Beyond interaction rates with power-law dependence on the temperature, Ref.~\cite{Taule:2022jrz}
    also considers transiently efficient interactions motivated by models featuring neutrino decay
    and inverse decay.
    In these scenarios the interaction rate, rather than changing monotonically, increases to some
    maximum and subsequently decreases with time.
    For simplicity, we restrict our results to those of the form
    \cref{eqn:interaction-rate-general}.
}

The de-/recoupling transition modulates the magnitude of the damping effect on scales near
the horizon at that time, $k_\star \equiv \mathcal{H}(\tau_\star)$.
In terms of this scale, we may write
\begin{align}\label{eqn:interaction-rate-ito-kstar-astar}
    \partial_\tau \kappa_\nu(\tau)
    = - \frac{\lambda T_\nu(\tau_\star)^n}{\left[ a(\tau) / a(\tau_\star) \right]^{n-1}}
    &= - k_\star
        \left( \frac{a(\tau)}{a(\tau_\star)} \right)^{1 - n},
\end{align}
assuming that the neutrino temperature decays as $1 / a(\tau)$ at all times.
We can replace the dependence on $a(\tau_\star)$ with $k / k_\star$ as follows.
First write the Friedmann equation in terms of the neutrino temperature and
fraction $\fnu(\tau)$ [given by \cref{eqn:f-nu-vs-tau}] as
\begin{align}
    \mathcal{H}(\tau)
    = \sqrt{ \frac{a(\tau)^2}{3 \Mpl^2} \frac{\bar{\rho}_\nu(\tau)}{\fnu(\tau)}}
    &= \sqrt{ \frac{7 \pi^2}{120 \fnu(\tau) \Mpl^2} }
        \frac{T_\nu(\tau_\star)^2}{a(\tau) / a(\tau_\star)},
\end{align}
with six effective neutrino degrees of freedom.
Taking the ratio of the above evaluated at $\tau$ and $\tau_\star$ leads to
\begin{align}\label{eqn:a-tau-over-a-tau-star}
    \frac{a(\tau)}{a(\tau_\star)}
    &= \sqrt{\frac{\fnu(\tau_\star)}{\fnu(\tau)}}
        \frac{k_\star}{\mathcal{H}(\tau)}.
\end{align}
Evaluating \cref{eqn:a-tau-over-a-tau-star} at $\tau = \tau_k \equiv 1/k$ and combining the result
with \cref{eqn:interaction-rate-ito-kstar-astar} yields
\begin{align}
    \partial_\tau \kappa_\nu(\tau)
    &= - k \left( \frac{a(\tau)}{a(\tau_k)} \right)^{1 - n}
        \left( \frac{k}{k_\star} \right)^{n - 2}
        \left( \frac{\fnu(\tau_k)}{\fnu(\tau_\star)} \right)^{\frac{n-1}{2}}.
\end{align}
When horizon crossing ($\tau_k$) and de-/recoupling ($\tau_\star$) are both sufficiently early
compared to equality ($\tau_\mathrm{eq})$ but still after weak decoupling and $e^+$-$e^-$
annihilation, the ratio of the neutrino abundances is unity.
In a radiation Universe,
\begin{align}
    \partial_\tau \kappa_\nu(\tau)
    &= - k \left( k \tau \right)^{1 - n}
        \left( \frac{k}{k_\star} \right)^{n - 2}.
\end{align}
Choosing instead to use \cref{eqn:a-tau-over-a-tau-star} with $\tau = \tau_\mathrm{eq}$ yields
\begin{align}
    \partial_\tau \kappa_\nu(\tau)
    &= - k_\mathrm{eq} \left( \frac{a(\tau)}{a(\tau_\mathrm{eq})} \right)^{1 - n}
        \left( \frac{k_\mathrm{eq}}{k_\star} \right)^{n - 2}
        \left( \frac{\fnu(\tau_\mathrm{eq})}{\fnu(\tau_\star)} \right)^{\frac{n-1}{2}}.
\end{align}
Note that $\fnu(\tau) / \fnu(\tau_\mathrm{eq}) \approx 2$ at times $\tau$ deep in the radiation
era.
Lastly, using
\begin{align}
    \kappa(\tau_1, \tau_2)
    &\equiv \int_{\tau_1}^{\tau_2} \ud \tilde{\tau} \,
            \partial_{\tilde{\tau}} \kappa_\nu(\tilde{\tau})
\end{align}
as shorthand notation, we define the opacity function
\begin{align}\label{eqn:def-opacity-function}
    \mathcal{O}_\nu(\tau_1, \tau_2)
    &= 1
        - e^{\kappa(\tau_1, \tau_2)},
\end{align}
which characterizes the probability that a single particle free streams between times $\tau_1$ and
$\tau_2$.

\subsection{Integral solutions}\label{sec:integral-solutions}

Before obtaining numerical results, we turn to formal, integral solutions for approximate, analytic
solutions and intuition.
Such integral equations were derived in Ref.~\cite{Stewart:1972a} for scalar, vector, and tensor
perturbations sourced by collisionless particles.
Ref.~\cite{Stewart:1972a} also outlined the procedure to obtain analytic solutions in series of
spherical Bessel functions (eventually further developed by Refs.~\cite{Dicus:2005rh,
Shchedrin:2012sp, Stefanek:2012hj}), anticipating that free-streaming matter would damp metric
perturbations.
Ref.~\cite{Rebhan:1991rt} (and also Ref.~\cite{Nachbagauer:1995ec}) employed an alternative power
series solution and appears to have been the first to present results for tensor perturbations,
finding a damping in amplitude of order $0.6$ for free-streaming--only solutions compared to
fluid-only ones, consistent with results here and elsewhere.
Refs.~\cite{Weinberg:2003ur,Weinberg:2008zzc} derived and applied this formalism to the neutrino
content of the Universe (after decoupling of the weak interactions) in the standard Big Bang model.

Refs.~\cite{Stewart:1972a,Rebhan:1991rt,Nachbagauer:1995ec,Weinberg:2003ur} (and subsequent work)
recast the coupled system of equations \cref{eqn:eoms-ito-x} into a single integro-differential
equation, exchanging the anisotropic stress's dependence on the distribution function for an
integral over the history of the gravitational wave itself.
To obtain this result, return to the momentum-integrated Boltzmann equation (before decomposing into
partial waves), \cref{eqn:ddeltafoverf-dtau-tensor}, which with
\cref{eqn:def-collision-perturbation-and-hierarchy} defining the collision term takes the form
\begin{align}\label{eqn:eom-F-lambda-not-hierarchy}
    0
    &= \partial_x F_\lambda^{(T)}(x, \mu)
        + i \mu F_\lambda^{(T)}(x, \mu)
        - \partial_x \kappa_\nu(x) F_\lambda^{(T)}(x, \mu)
        + 2 \partial_x \hD_\lambda(x).
\end{align}
Observe that \cref{eqn:eom-F-lambda-not-hierarchy} has the formal solution (from some initial
condition at $x_i$)
\begin{align}\label{eqn:formal-integral-solution-F-lambda}
    F_\lambda^{(T)}(x, \mu)
    &= e^{
            - i \mu (x - x_i)
            + \kappa_\nu(x_i, x)
        }
        F_\lambda^{(T)}(x_i, \mu)
        - 2 \int_{x_i}^{x} \ud u \,
        e^{\kappa_\nu(u, x)}
        e^{-i \mu (x - u)} \partial_u \hD_\lambda(u).
\end{align}
Replacing the exponential inside the integral with the partial wave expansion of a plane wave and
recomputing the anisotropic stress [\cref{eqn:anisotropic-stress-ito-moments}], the gravitational
wave equation of motion becomes
\begin{align}\label{eqn:integral-equation}
    \partial_x^2 \hD_\lambda
        + 2 \frac{\partial_x a}{a} \partial_x \hD_\lambda
        + \hD_\lambda
    &= - 24 \fnu(\tau)
        \left( \frac{\partial_x a}{a} \right)^2
        \int_{x_i}^{x} \ud u \,
        \left[ 1 - \mathcal{O}_\nu(u, x) \right]
        K(x - u) \partial_u \hD_\lambda,
\end{align}
where we take the initial condition $F_\lambda^{(T)}(x_i, \mu) = 0$ and define
\begin{align}
    K(y)
    &\equiv \frac{1}{15} j_0(y)
        + \frac{2}{21} j_2(y)
        + \frac{1}{35} j_4(y)
    = \frac{j_2(y)}{y^2}
\end{align}
in terms of the spherical Bessel functions of the first kind $j_\gamma(x)$.
\Cref{eqn:integral-equation} reproduces the form obtained in Ref~\cite{Weinberg:2003ur} for the
collisionless case, $\mathcal{O}_\nu(u, x) = 0$.
Integro-differential equations like \cref{eqn:integral-equation} may be solved numerically via an
iterative algorithm or by discretizing in time and solving the resulting linear system.
Alternatively, \cref{eqn:integral-equation} may be solved analytically in a series of spherical
Bessel functions, as applied to the case of inflationary gravitational waves (and collisionless
neutrinos) in Refs.~\cite{Dicus:2005rh,Shchedrin:2012sp,Stefanek:2012hj}.

Though the source term in the integro-differential form \cref{eqn:integral-equation} depends upon
$\hD_\lambda$ itself, it proves useful nonetheless to study the formal solution in terms of Green
functions.
First, define the rescaled tensor perturbation $v_\lambda \equiv a \hD_\lambda$, which does
not decay with the expansion of the Universe.
The gravitational wave equation of motion \cref{eqn:h-lambda-eom} in Fourier space takes the form
\begin{align}\label{eqn:tensor-eom-rescaled}
    v_\lambda''(\tau, \mathbf{k})
        + \left( k^2 - \frac{a''}{a} \right) v_\lambda(\tau, \mathbf{k})
    &= a \frac{2 a^2}{\Mpl^2} \pi_\lambda^T(\tau, \mathbf{k}).
\end{align}
For simplicity, we here fix a single-component Universe [\cref{eqn:single-component-solution}] at the
background level and again work in terms of the coordinate $x \equiv k \tau$.
The particular solution to \cref{eqn:tensor-eom-rescaled} is a convolution of the source and the Green
function $G(x; \tilde{x})$ that satisfies
\begin{align}\label{eqn:green-function-eom}
    \partial_x^2 G(x, \tilde{x})
    + \left( 1 - \frac{\alpha (\alpha - 1)}{x^2} \right) G(x, \tilde{x})
    = \delta(x - \tilde{x}).
\end{align}
The causal solution is
\begin{align}
    G(x; \tilde{x})
    &= \Theta(x - \tilde{x})
        x \tilde{x} \left[
            j_{\alpha-1}(\tilde{x}) y_{\alpha-1}(x)
            - j_{\alpha-1}(x) y_{\alpha-1}(\tilde{x})
        \right],
\end{align}
where $\Theta$ is the Heaviside function and $j_\gamma$ and $y_\gamma$ are the order-$\gamma$
spherical Bessel functions of the first and second kind, respectively.
Denoting the homogeneous solution as $\hD_\lambda^{(0)}(x)$ and substituting
$a(\tau) = (k \tau)^\alpha$, the full (homogeneous plus particular) solution reads
\begin{align}
    \hD_\lambda(x)
    = \hD_\lambda^{(0)}(x)
        + \frac{1}{x^\alpha}
        \int_{x_i}^{x} \ud \tilde{x} \,
        G(x; \tilde{x})
        \tilde{x}^\alpha
        \frac{2 \tilde{x}^{2 \alpha}}{k^2 \Mpl^2} \pi_\lambda^T(\tilde{x}).
\end{align}
We are mainly interested in the radiation era, in which $\alpha = 1$ and the Green function is
\begin{align}\label{eqn:green-function-radiation}
    G(x; \tilde{x})
    &= \sin x \cos \tilde{x}
        - \sin \tilde{x} \cos x
    = \sin (x - \tilde{x}).
\end{align}
Finally, plugging in the integral solution for $\pi_\lambda^T$ as in \cref{eqn:integral-equation}
and taking $\fnu$ constant yields
\begin{align}\label{eqn:green-function-soln-radiation}
    \hD_\lambda(x)
    &= \hD_\lambda^{(0)}(x)
        - \frac{24 \fnu}{x}
        \int_{x_i}^x \ud \tilde{x} \,
        \frac{\sin\left( x - \tilde{x} \right)}{\tilde{x}}
        \int_{x_i}^{\tilde{x}} \ud u \,
        \left[ 1 - \mathcal{O}_\nu(u, \tilde{x}) \right]
        K(\tilde{x} - u)
        \partial_u \hD_\lambda(u).
\end{align}
Since $K(y) \approx - \sin(y) / y^3$ for $y \gg 1$, at late times the integral over $u$ receives
most of its support where $u \sim \tilde{x}$.
The contributions to the inhomogeneous term from late times (relative to horizon crossing) are
therefore suppressed, localizing the importance of anisotropic stress near horizon crossing.
As such, stochastic backgrounds of gravitational waves encode the interaction history of
relativistic species.
Though \cref{eqn:green-function-soln-radiation} does not permit a solution by direct integration, we
might hope to gain insight by approximating $\partial_u \hD_\lambda(u)$ with the homogeneous
solution (i.e., to zeroth order in an expansion in $\fnu$).
We explore the utility of this approach in application to inflationary gravitational waves in
\cref{sec:results-inflationary-noninteracting}.

We close by connecting the above formalism to the observable gravitational wave spectrum.
We first phrase the formal solution in terms of a modulation of the amplitude and phase of
the solution as induced by free-streaming particles.
Decompose the solution \cref{eqn:green-function-soln-radiation} onto sine and cosine modes
[i.e., using the first form of \cref{eqn:green-function-radiation}] as
\begin{align}\label{eqn:h-solution-expanded-sin-cos}
    \hD_\lambda(x)
    &\equiv \hD_{\lambda, s}(x) \frac{\sin(x)}{x}
        + \hD_{\lambda, c}(x) \frac{\cos(x)}{x} \\
    &\equiv
        \left[
            \hD_{\lambda, s}
            + \mathcal{I}_s(x)
        \right]
        \frac{\sin(x)}{x}
        +
        \left[
            \hD_{\lambda, c}
            + \mathcal{I}_c(x)
        \right]
        \frac{\cos(x)}{x},
\end{align}
where $\hD_{\lambda, s}$ and $\hD_{\lambda, c}$ parameterize the homogeneous solution while the
inhomogeneous contributions are
\begin{align}\label{eqn:I-cs-defn}
    \begin{pmatrix}
        \mathcal{I}_s(x) \\ \mathcal{I}_c(x)
    \end{pmatrix}
    &\equiv
        24 \Omega_\nu
        \int_{x_i}^x \ud \tilde{x} \,
        \frac{1}{\tilde{x}}
        \begin{pmatrix}
            - \cos \tilde{x} \\ \sin \tilde{x}
        \end{pmatrix}
        \int_{x_i}^{\tilde{x}} \ud u \,
        \left[ 1 - \mathcal{O}_\nu(u, \tilde{x}) \right]
        K(\tilde{x} - u) \partial_u \hD_\lambda(u).
\end{align}
Comparing \cref{eqn:h-solution-expanded-sin-cos} to
\begin{align}\label{eqn:def-h-ito-amp-and-phase}
    \hD_\lambda(x)
    &\equiv A(x) \frac{\sin \left[ x + \varphi(x) \right]}{x}
\end{align}
reveals that the (time-dependent) amplitude and phase of the solution are respectively modulated by
free-streaming particles as
\begin{align}
    A(x)
    &= \sqrt{\hD_{\lambda, s}(x)^2 + \hD_{\lambda, c}(x)^2}
    \label{eqn:amplitude-ito-hs-hc}
    \\
    \varphi(x)
    &= \arctan \left( \frac{\hD_{\lambda, c}(x)}{\hD_{\lambda, s}(x)} \right).
    \label{eqn:phase-ito-hs-hc}
\end{align}
Parametrizing in terms of an amplitude and phase is valid at all times only in a radiation-dominated
Universe.
In any single-component Universe, however, \cref{eqn:def-h-ito-amp-and-phase} is still valid
asymptotically, since the late-time limits of spherical Bessel functions of any order are
proportional to sine and cosine.
We therefore define $A_\infty \equiv A(\infty)$ and $\varphi_\infty \equiv \varphi(\infty)$.
[Note as well that, in terms of the primordial initial condition
$\hD_{\lambda, 0} = \lim_{k \tau \to 0} \hD_\lambda(k \tau)$,
$A_{\infty}[f_\mathrm{fs} = 0] = \hD_{\lambda, 0}$ only in a radiation dominated
Universe; for general $\alpha$ the value is
$2^\alpha \Gamma(\alpha + 1/2) \hD_{\lambda, 0} / \sqrt{\pi}$.]

We present main results in terms of the relative change to the spectral energy density in
gravitational waves [\cref{eqn:omega-gw-spectrum-ito-power-spectrum}],
\begin{align}
    \frac{\Omega_\mathrm{GW}}{\Omega_\mathrm{GW}[f_\mathrm{fs} = 0]}
    &= \left( \frac{A_\infty }{ A_{\infty}[f_\mathrm{fs} = 0] } \right)^2,
\end{align}
compared to a Universe with a free-streaming fraction $f_\mathrm{fs} = 0$ (and therefore no
anisotropic stress).
The phase shift is likely unobservable for direct detection of the strain of stochastic backgrounds,
which average signals over an integration time much longer than the gravitational-wave period.
However, phase shifts would impact the spectrum of CMB polarization, since the visibility function
for photons is sharply peaked at recombination (making the CMB an effective snapshot of that
moment in time).

\section{Damping of inflationary gravitational waves}\label{sec:inflation}

We now apply the formalism established in \cref{sec:background} to develop a semianalytic
understanding of the effect of relativistic particles on the primordial gravitational wave
background generated during inflation.
Using the formal results of \cref{sec:integral-solutions}, in
\cref{sec:results-inflationary-noninteracting} we study the amplitude modulation and phase shift of
gravitational waves due to free-streaming particles.
Building on intuition from these semianalytic results in noninteracting scenarios,
\cref{sec:results-inflation} presents numerical results for a variety of scenarios in which
interactions decouple or recouple, focusing on the effect of SM neutrinos.

\subsection{Results in noninteracting scenarios}\label{sec:results-inflationary-noninteracting}

Gravitational waves generated during inflation exit the horizon in the asymptotic past and remain
frozen until their wavelength reenters the horizon.
This case corresponds to an initial condition far outside the horizon
($x = k \tau = k / \mathcal{H} \ll 1$) of
\begin{subequations}\label{eqn:inflationary-initial-condition}
\begin{align}
    \lim_{x \to 0} \hD_\lambda(x)
    &= \hD_{\lambda, 0} \\
    \lim_{x \to 0}  \partial_x \hD_{\lambda, 0}(x)
    &= 0 \\
    \lim_{x \to 0} F^{(T)}_{\lambda, l}(x)
    &= 0
\end{align}
\end{subequations}
in terms of the primordial amplitude $\hD_{\lambda, 0}$.
The homogeneous solution (i.e., in the absence of anisotropic stress) to \cref{eqn:h-lambda-eom} is
$\hD_\lambda^{(0)}(x) = \hD_{\lambda, 0} j_{0}(x) = \hD_{\lambda, 0} \sin x / x$.
In the language of \cref{eqn:h-solution-expanded-sin-cos}, $\hD_{\lambda, s} = \hD_{\lambda, 0}$
and $\hD_{\lambda, c} = 0$.

\begin{figure}[t]
    \centering
    \includegraphics[width=\textwidth]{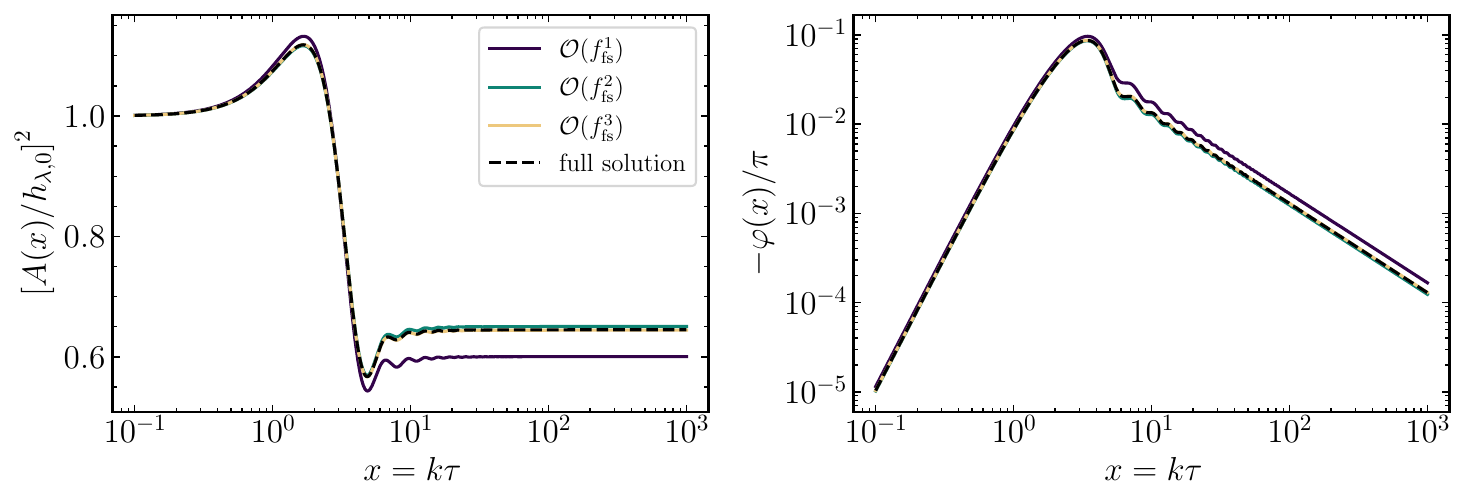}
    \caption{
        Time dependent amplitude and phase, \cref{eqn:amplitude-ito-hs-hc,eqn:phase-ito-hs-hc}, in
        the presence of free-streaming radiation.
        The solid curves depict the semianalytic results at varying orders in $f_\mathrm{fs}$,
        while the black dashed curve shows the quantities evaluated with a full solution
        to the Boltzmann hierarchy.
        All results fix the free-streaming fraction to that of SM neutrinos,
        $f_\mathrm{fs} = \Rnu = 0.40523$.
    }
    \label{fig:I-sc-amp-phase}
\end{figure}

As argued in \cref{sec:integral-solutions}, the impact of free-streaming particles decays with time.
In practice, this means that $A(x)$ and $\varphi(x)$ converge\footnote{
    Within the present framework, there is no means by which, well after horizon crossing, the tensor
    anisotropic stress $\pi^T_\lambda$ can grow with time (or remain constant), so the integrals are
    convergent.
    The tensor anisotropic stress would grow, if, for example, another species decays into
    relativistic particles or directly sources gravitational waves.
} to their late-time values $A_\infty$ and $\varphi_\infty$ relatively quickly; in
\cref{app:semianalytic-inflation} we show that, at late times $x \gg 1$,
\begin{subequations}\label{eqn:amplitude-phase-analytic}
\begin{align}
    \left( \frac{A(x)}{\hD_{\lambda, 0}} \right)^2
    &\approx
        \left( 1 - \frac{5}{9} f_\mathrm{fs} \right)^2
            + \frac{
                f_\mathrm{fs}^2
                - 2 f_\mathrm{fs} \left( 1 - 5 f_\mathrm{fs} / 9 \right) \sin^2 x
            }{x^2}
    \label{eqn:amplitude-sq-analytic}
    \\
    \varphi(x)
    &\approx - \frac{1}{x} \frac{f_\mathrm{fs}}{1 - 5 f_\mathrm{fs} / 9}
    \label{eqn:phase-analytic}
\end{align}
\end{subequations}
when inserting the homogeneous solution $\hD_{\lambda}^{(0)}$ into the integrals in
\cref{eqn:I-cs-defn} (and replacing $\Omega_\nu$ with $f_\mathrm{fs}$).
We display the full result for all times $x$ in \cref{fig:I-sc-amp-phase}, which shows for
comparison the result from a full numerical solution to the Boltzmann hierarchy (as described in
\cref{app:numerical-implementation}).\footnote{
    \Cref{fig:I-sc-amp-phase} also displays results at higher orders in the expansion in
    $f_\mathrm{fs}$, i.e., those obtained by successively inserting solutions into
    \cref{eqn:green-function-soln-radiation} and computing integrals by numerical quadrature.
    The iterative procedure quickly converges to the fully numerical result.
    A similar approach was recently applied to scalar perturbations in
    Refs.~\cite{Kamionkowski:2021njk,Ji:2022iji}.
}
While the semianalytic results for $\mathcal{I}_s$ and $\mathcal{I}_c$ are simply proportional to
$f_\mathrm{fs}$, the amplitude and phase depend nonlinearly on $f_\mathrm{fs}$, for which reason
\cref{fig:I-sc-amp-phase} fixes $f_\mathrm{fs} = \Rnu$ for SM neutrinos after weak decoupling.
As one might anticipate by the form of \cref{eqn:amplitude-ito-hs-hc}, the overall amplitude of
gravitational waves is in fact enhanced just after horizon crossing due to the sourcing of the
cosine mode (which is not present in the absence of anisotropic stress).
The sine mode is initially enhanced as well, but $\mathcal{I}_s(x)$ begins decreasing rapidly from
its maximum just before $\abs{\mathcal{I}_c}$ (and also the phase shift) is maximized.
The amplitude rapidly drops during the first oscillation and then begins to oscillate as
$\sin^2 x / x^2$ about its asymptotic value, as per \cref{eqn:amplitude-sq-analytic}.
At the same time, the phase decays linearly as in \cref{eqn:phase-analytic}.
See Ref.~\cite{Bashinsky:2005tv} for an alternative derivation of these results using real-space
Green function methods.

Finally, we comment on the damping factor's dependence on $f_\mathrm{fs}$.
The aforementioned analytic, series solution to
\cref{eqn:integral-equation}~\cite{Dicus:2005rh,Shchedrin:2012sp,Stefanek:2012hj} was employed in
Ref.~\cite{Boyle:2005se} to compute an analytic approximation to the damping factor for inflationary
gravitational waves that is accurate to $0.1 \%$ for all $f_\mathrm{fs}$ between 0 and 1.
Our numerical solutions to \cref{eqn:eoms-ito-x} [with the initial condition
\cref{eqn:inflationary-initial-condition}] verify the accuracy of the analytic approximation of
Ref.~\cite{Boyle:2005se} over the full range of $f_\mathrm{fs}$ between $0$ and $1$.
However, by numerological coincidence, the (far simpler) fitting function
\begin{align}\label{eqn:damping-factor-fitting}
    \frac{A_\infty}{A_\infty[f_\mathrm{fs} = 0]}
    &= \left(
            1 + \frac{f_\mathrm{fs}}{9} + \frac{f_\mathrm{fs}^2}{23}
        \right)
        e^{- 2 f_\mathrm{fs} / 3}
\end{align}
reproduces numerical results to one part in $10^4$ for all $f_\mathrm{fs}$.
This fitting formula, to leading order in $f_\mathrm{fs}$, agrees with the semianalytic result of
(the square root of) \cref{eqn:amplitude-sq-analytic}, $1 - 5 f_\mathrm{fs} / 9$.

\subsection{Results in interacting scenarios}\label{sec:results-inflation}

Having reviewed the known effects of free-streaming particles on gravitational waves, we turn to
more general scenarios in which relativistic species transition between weakly and strongly
interacting states.
We first establish expectations based upon the semianalytic results of
\cref{sec:results-inflationary-noninteracting}, inferring from \cref{eqn:amplitude-phase-analytic} and
\cref{fig:I-sc-amp-phase} how the instantaneous decoupling or recoupling of interactions would
modulate the gravitational wave power spectrum.

Consider first dark radiation that decouples from interactions.
For wavenumbers $k$ much smaller than the horizon size at the transition $k_\star$, decoupling
occurs at $x_\star = k \tau_\star \ll 1$, well before horizon crossing; these modes therefore
experience the full damping effect.
In the opposite case, free streaming is postponed until well after horizon crossing---beyond when
the bulk of the damping effect would take place.
A residual phase shift and amplitude suppression occur respectively at linear and quadratic order
in $k_\star / k$, evident by inspecting the integrands of \cref{eqn:I-cs-defn}.
For modes entering the horizon near the time of decoupling, the results are more complex.
For example, some modes do not experience the initial enhancement of the cosine mode (and the
overall amplitude), and could therefore be \textit{more} suppressed than modes that enter the
horizon well after decoupling.
The decaying oscillations in time in \cref{eqn:amplitude-sq-analytic} would imprint a relic
amplitude oscillation (in $k$) for modes that begin oscillating slightly sooner before decoupling.
Indeed, this exact oscillatory effect for neutrino decoupling from the weak interactions was
observed in Refs.~\cite{Watanabe:2006qe,Saikawa:2018rcs,Kite:2021yoe} and (correctly) attributed to
use of the instantaneous decoupling approximation.
In reality, decoupling (for interaction rates proportional to, e.g., $T^5$) takes a nonvanishing
amount of time, smearing out such features; however, they may be evident in scenarios where
decoupling occurs due to a rapid drop in number density, such as dark hydrogen recombination.

A similar analysis of \cref{fig:I-sc-amp-phase} for recoupling scenarios suggests that modes that
reenter the horizon near the time of recoupling would be slightly \textit{enhanced}, since
interactions would rapidly suppress anisotropic stress before the bulk of the amplitude suppression
occurs.
These modes would also retain a nonzero phase shift.
The effect for modes with wavenumbers larger than $k_\star$ would be truncated: the amplitude is
damped by an increasing amount with increasing $k$, saturating at the value for the noninteracting
case, and the late-time phase shift approaches zero linearly with $k_\star / k$.

Expectations in hand, we now present numerical solutions to \cref{eqn:eoms-ito-x}.
Rather than employing an iterative method to solve the integro-differential form of the system,
we find it simple and efficient enough to solve the original system of equations [i.e.,
\cref{eqn:eoms-ito-x} for the gravitational waves and the Boltzmann hierarchy] directly using
standard methods for ordinary differential equations.
We outline our numerical implementation in \cref{app:numerical-implementation}.

\begin{figure}[t!]
    \centering
    \includegraphics[width=\textwidth]{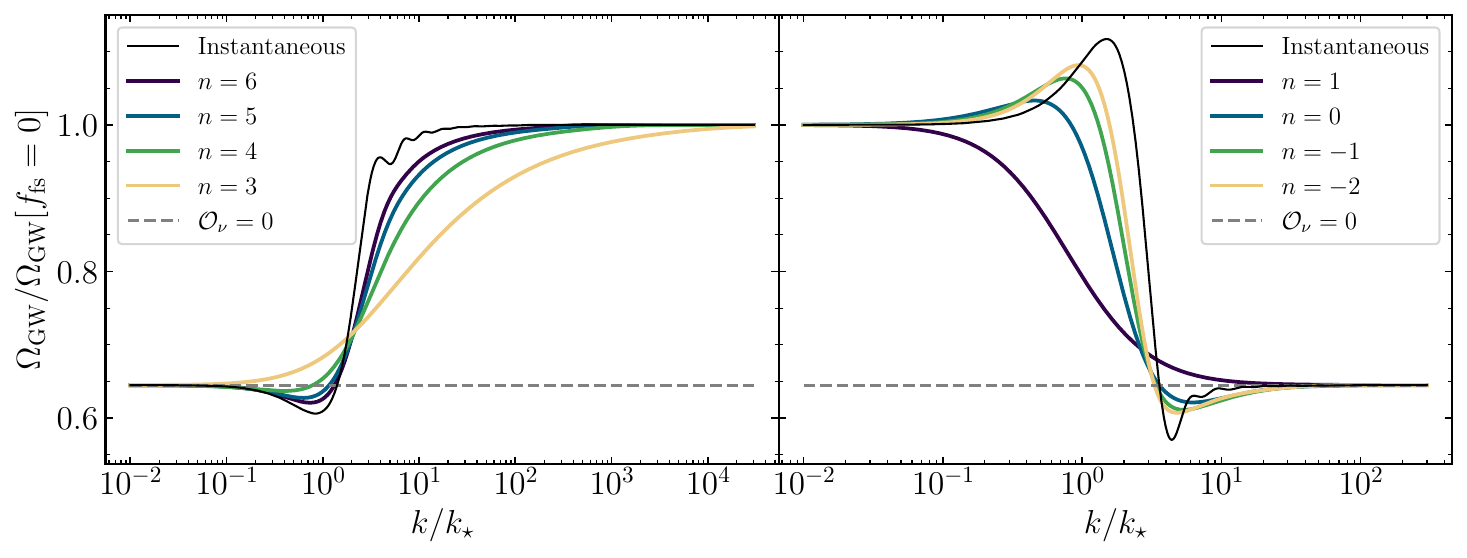}
    \caption{
        Damping factor (in a radiation Universe) for inflationary gravitational waves in decoupling
        (left) and recoupling (right) scenarios with interaction rates of varying temperature
        dependence $T^n$ [\cref{eqn:interaction-rate-general}] as denoted by the legends (colored
        curves).
        The spectrum is normalized relative to that for a Universe with no free-streaming particles,
        and $k_\star$ denotes the horizon size at the time of de-/recoupling.
        The dashed grey curve depicts the result in the noninteracting case (i.e., zero opacity at
        all times), and the thin black curve that for instantaneous de-/recoupling.
        All results fix the abundance of the radiation sector to that of SM neutrinos,
        $\fnu = \Rnu = 0.40523$.
    }
    \label{fig:damping-vs-no-nu-vary-n-rad}
\end{figure}

We first consider a radiation Universe (a suitable approximation for de-/recoupling occurring
sufficiently early before matter-radiation equality, i.e., $k_\star \gg k_\mathrm{eq}$).
\Cref{fig:damping-vs-no-nu-vary-n-rad} displays the amplitude modulation relative to that for a
Universe with no free-streaming radiation ($f_\mathrm{fs} = 0$) for interaction rates with varying
temperature dependence $T^n$.
For comparison, \cref{fig:damping-vs-no-nu-vary-n-rad} includes curves representing the
instantaneous de-/recoupling cases (approximated with $n = \pm 25$ solutions for numerical
convenience), which exhibit the oscillatory feature anticipated from \cref{fig:I-sc-amp-phase} (and
observed in, e.g., Refs.~\cite{Watanabe:2006qe,Saikawa:2018rcs,Kite:2021yoe}).
(Note that the $n = 5$ curve corresponds to the result for standard neutrino decoupling, i.e., when
the weak interactions become inefficient.)
Results in all cases exhibit the nearly sigmoidal behavior expected from the preceding discussion,
centered on some scale near $k_\star$.
Additionally, in scenarios where de-/recoupling occurs rapidly enough ($\abs{n - 2} > 1$), we also
observe the anticipated overshoot: modes which enter the horizon just after decoupling (or before
recoupling) are damped more than in the noninteracting case.
Furthermore, in the recoupling scenario modes with wavenumber close to $k_\star$ are indeed
enhanced via the sourcing of the cosine mode.

\begin{figure}[t!]
    \centering
    \includegraphics[width=\textwidth]{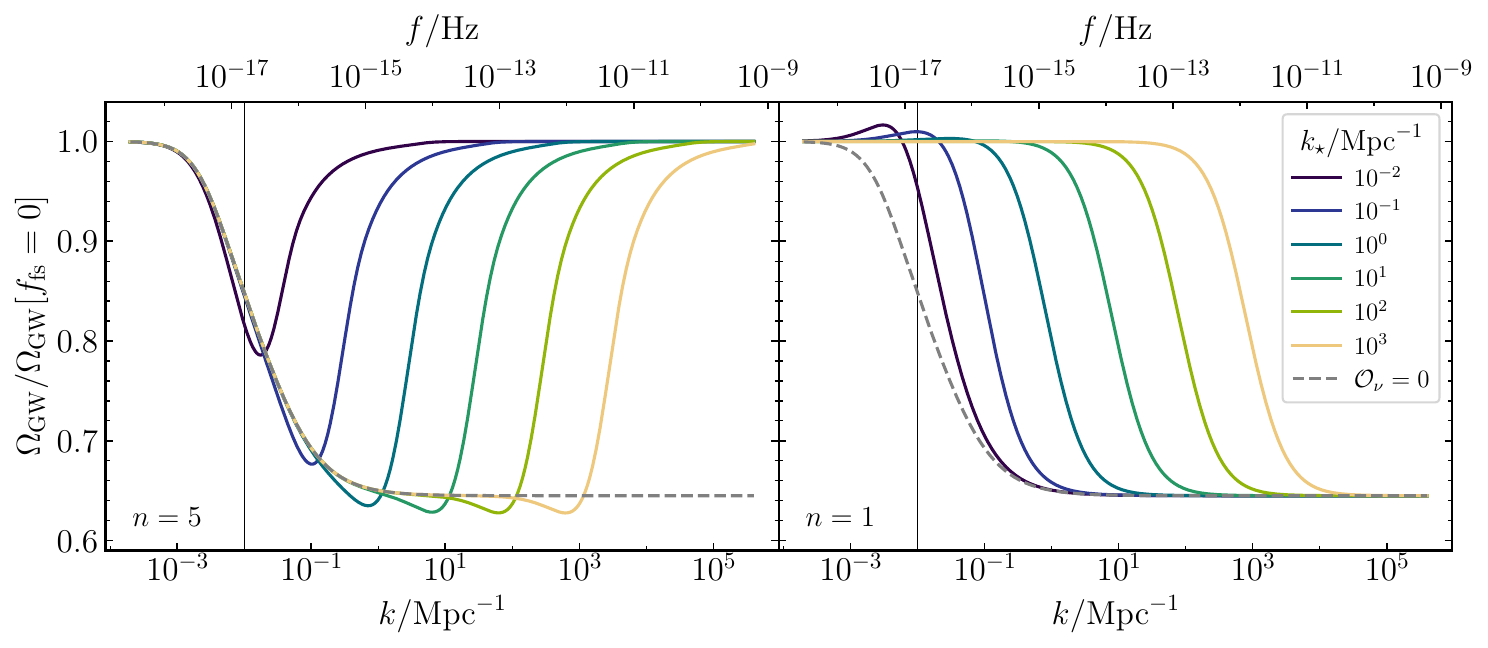}
        \caption{
        Impact of varying the epoch of de-/recoupling on the amplitude of inflationary gravitational
        waves.
        The damping factor for interactions mediated by heavy ($n = 5$, left) and light ($n = 1$,
        right) degrees of freedom are plotted for various coupling strengths parameterized in terms
        of the horizon scale $k_\star$ at the time of de-/recoupling, denoted by the legend.
        The dashed grey line depicts the result for neutrinos with no beyond-the-SM interactions.
        Vertical lines mark the horizon scale at matter-radiation equality; the abundance of
        neutrinos, and therefore the damping of gravitational waves, decays as matter comes
        to dominate the Universe's energy budget.
        All results fix the abundance of the radiation sector to that of SM neutrinos,
        $\fnu = \Rnu = 0.40523$.
    }
    \label{fig:damping-heavy-light}
\end{figure}

To study scenarios relevant to the CMB, we adopt a matter-radiation Universe and consider $n = 5$
and $1$ as benchmark decoupling and recoupling cases.
\Cref{fig:damping-heavy-light} displays results for scenarios of varying coupling strengths (phrased
as usual in terms of the horizon size at de-/recoupling in units of $\mathrm{Mpc}^{-1}$.)
For the decoupling case, the interplay between the transition to free streaming and the onset of
matter domination [which reduces the neutrino abundance per \cref{eqn:f-nu-vs-tau}] introduces a
dip-like feature in the gravitational wave power spectrum.
In recoupling scenarios, the reintroduction of interactions introduces a similar feature as the
transition to matter domination, but at smaller scales.
When recoupling occurs near matter-radiation equality, the two effects conspire to mildly enhance
the gravitational wave spectrum on scales slightly larger than the horizon at equality.

\begin{figure}[t!]
    \centering
    \includegraphics[width=\textwidth]{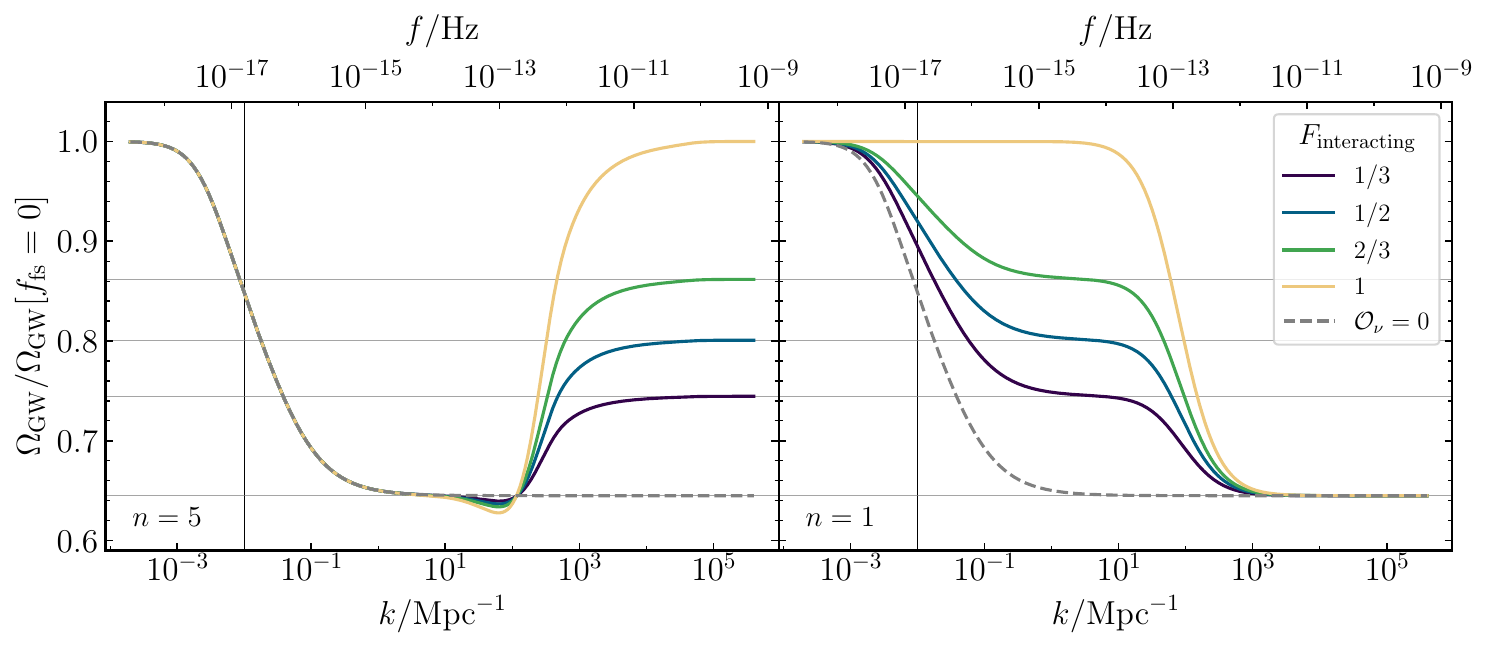}
    \caption{
        Damping factor for inflationary gravitational waves as a function of scale when a differing
        number of the neutrino eigenstates participate in interactions mediated by heavy ($n = 5$,
        left) and light ($n = 1$, right) degrees of freedom.
        All results take the horizon scale at the time of de-/recoupling to be
        $k_\star = 10^{2} \, \mathrm{Mpc}^{-1}$ and fix $\fnu$ to the Standard Model value,
        $\Rnu = 0.40523$, of which only a fraction $F_\mathrm{interacting}$ (labeled by the legend)
        comprises interacting species.
        A curve assuming that effectively half of the neutrinos interact is included for
        illustrative purposes.
        Horizontal lines indicate the squared damping factors [evaluated with the fitting function
        \cref{eqn:damping-factor-fitting}] for each case when taking
        $f_\mathrm{fs} = F_\mathrm{interacting} \fnu$.
        The vertical lines indicate the horizon scale at matter-radiation equality.
    }
    \label{fig:damping-heavy-light-partial}
\end{figure}

Finally, the above results extend straightforwardly if only some fraction $F_\mathrm{interacting}$
of the relativistic species self-interacts.
In light of CMB constraints on the amount of energy in radiation (i.e., neutrinos) that can be
fluidlike~\cite{Baumann:2015rya,Brust:2017nmv,Blinov:2020hmc} as well as laboratory constraints on
neutrino self-interactions (see, e.g., Ref.~\cite{Blinov:2019gcj}), such scenarios are more probable
(if neutrino interactions remain important during the formation of the CMB\footnote{
    However, \textit{Planck} and ACT polarization data appear to drive differing preferences for
    strongly interacting neutrinos; see Ref.~\cite{Kreisch:2022zxp} for a recent investigation.
    \textit{Planck} constraints on strongly interacting neutrinos are also weaker when only one or
    two eigenstates interact~\cite{Das:2020xke,Brinckmann:2020bcn}.
}).
\Cref{fig:damping-heavy-light-partial} displays results for benchmark decoupling and recoupling
scenarios, demonstrating that the damping before decoupling (or after recoupling) corresponds to
the expectation for a Universe with a fraction $(1 - F_\mathrm{interacting}) \fnu$ of energy
in free-streaming, relativistic particles.
Analogous results would apply to dark radiation sectors, but the effect would be proportionally
smaller (given cosmological constraints on new light relics).

\section{Damping of causal gravitational waves}\label{sec:causal}

In contrast to inflation, which sources gravitational waves that remain frozen until they eventually
reenter the horizon, most other cosmological sources of stochastic backgrounds (such as phase
transitions or resonant particle production) occur sufficiently far inside the horizon that the
effect of free-streaming particles is unimportant.
On the other hand, in these scenarios the gravitational wave source is typically quadratic in some
degree of freedom carrying anisotropic stress (e.g., the velocity of a fluid, the gradient of a
scalar field, or electric and magnetic fields), coupling long-wavelength gravitational waves to the
subhorizon, ``causal'' dynamics of the source.
Ref.~\cite{Hook:2020phx} provided a model-independent description of the infrared,
``causality-limited'' gravitational wave spectrum from causal processes in general cosmologies; we
now briefly review this formalism, concentrating on modes that were outside the horizon during the
phase transition.\footnote{
    Though the treatment is agnostic to the nature of the subhorizon physics at play, we follow
    Ref.~\cite{Hook:2020phx} in simply referring to it as a phase transition.
}
We focus on the effect of free-streaming radiation studied in Ref.~\cite{Hook:2020phx} and then
extend results to interacting scenarios as studied in the previous section.

\subsection{Results in noninteracting scenarios}\label{sec:results-causal-noninteracting}

On scales sufficiently larger than those that are dynamical during the phase transition, the
anisotropic stress tensor (in Fourier space) is independent of wavenumber
$k$~\cite{Hook:2020phx,Caprini:2009fx}.
Furthermore, for gravitational waves with such long characteristic timescales, the phase transition
essentially occurs instantaneously.
Such superhorizon tensor modes are well described by an initial condition wherein the gravitational
wave is spontaneously ``kicked'' at the time $\tau_i$ of the phase transition:
\begin{subequations}\label{eqn:causal-initial-condition}
\begin{align}
    \hD_\lambda(\tau_i, \mathbf{k})
    &= 0 \\
    \partial_\tau \hD_\lambda(\tau_i, \mathbf{k})
    &= J_i \\
    F^{(T)}_{\lambda, l}(\tau_i, \mathbf{k})
    &= 0.
\end{align}
\end{subequations}
In general, the wavenumber-independent magnitude of the source $J_i$ depends on the details of the
subhorizon source.
While these initial conditions (and the equations themselves, in a radiation Universe) are
independent of wavenumber, the resulting solutions are not: each mode spends a different amount of
time outside the horizon after being sourced.
As a result, the low-frequency tail of the gravitational wave spectrum can carry information about
the equation of state and free-streaming content of the Universe (and the time dependence thereof)
during this period.
Ref.~\cite{Hook:2020phx} showed that, in terms of the equation of state of the Universe $w$ and in
the absence of free-streaming particles, the spectrum (for wavenumbers that were superhorizon during
the phase transition) scales as
\begin{align}\label{eqn:omega-gw-causal-general-eos}
    \Omega_\mathrm{GW}(\tau, k)
    &\propto k^{(1 + 15 w) / (1 + 3 w)},
\end{align}
reproducing the well-known $k^3$ behavior in a radiation Universe.

The superhorizon modes in this case are (strongly) overdamped, sourced harmonic oscillators.
Soon after being sourced, the gravitational wave approaches a constant-amplitude solution with
$\hD_\lambda \approx J_i / \mathcal{H}(\tau_i)$ if its wavenumber is sufficiently smaller than
the horizon ($k \tau_i \lesssim 10^{-3/2}$)---otherwise, it begins oscillating before saturating
this maximal, frozen amplitude.
After this point they resemble inflationary gravitational waves, suggesting that the damping effect
of free-streaming radiation at horizon crossing is still relevant.
However, their initial velocity also sources anisotropic stress in free-streaming particles [evident
in the formal solution for $F_\lambda^{(T)}$, \cref{eqn:formal-integral-solution-F-lambda}].
As shown by Ref.~\cite{Hook:2020phx}, the damping effect of free-streaming particles is more
important just after $\hD_\lambda$ is ``kicked'' than after horizon crossing, leading to a
qualitatively different signature.
To understand the resulting dynamics outside the horizon, take the $k \tau \ll 1$ limit of the
integral equation, \cref{eqn:integral-equation}.
As $K(y) = 1/15 + \mathcal{O}(y^2)$, the gravitational wave then evolves according to\footnote{
    This limit is applicable to inflationary gravitational waves as discussed in
    \cref{sec:results-inflation} but is irrelevant: frozen-out gravitational waves have
    negligible velocity, so the right-hand side of \cref{eqn:integral-equation-superhorizon}
    is likewise small when well outside the horizon.
}
\begin{align}\label{eqn:integral-equation-superhorizon}
    \partial_x^2 \hD_\lambda
        + 2 \frac{\partial_x a}{a} \partial_x \hD_\lambda
        + \hD_\lambda
    &= - \frac{8}{5} f_\mathrm{fs}
        \left( \frac{\partial_x a}{a} \right)^2
        \left[
            \hD_\lambda(x)
            - \hD_\lambda(x_i)
        \right].
\end{align}
In radiation domination [\cref{eqn:single-component-solution}] with $f_\mathrm{fs}$ constant,
the solution to \cref{eqn:integral-equation-superhorizon} from the causal initial condition
\cref{eqn:causal-initial-condition} is given in terms of spherical Bessel functions as
\begin{align}\label{eqn:causal-sol}
    \hD_\lambda(x)
    &= \frac{J_i x_i^2}{k} \left[
            j_\gamma(x_i) y_\gamma(x)
            - j_\gamma(x) y_\gamma(x_i)
        \right],
\end{align}
where
\begin{align}
    \gamma
    &= \frac{-1 + \sqrt{1 - 32 f_\mathrm{fs} / 5}}{2}.
\end{align}

\begin{figure}[t!]
    \centering
    \includegraphics[width=\textwidth]{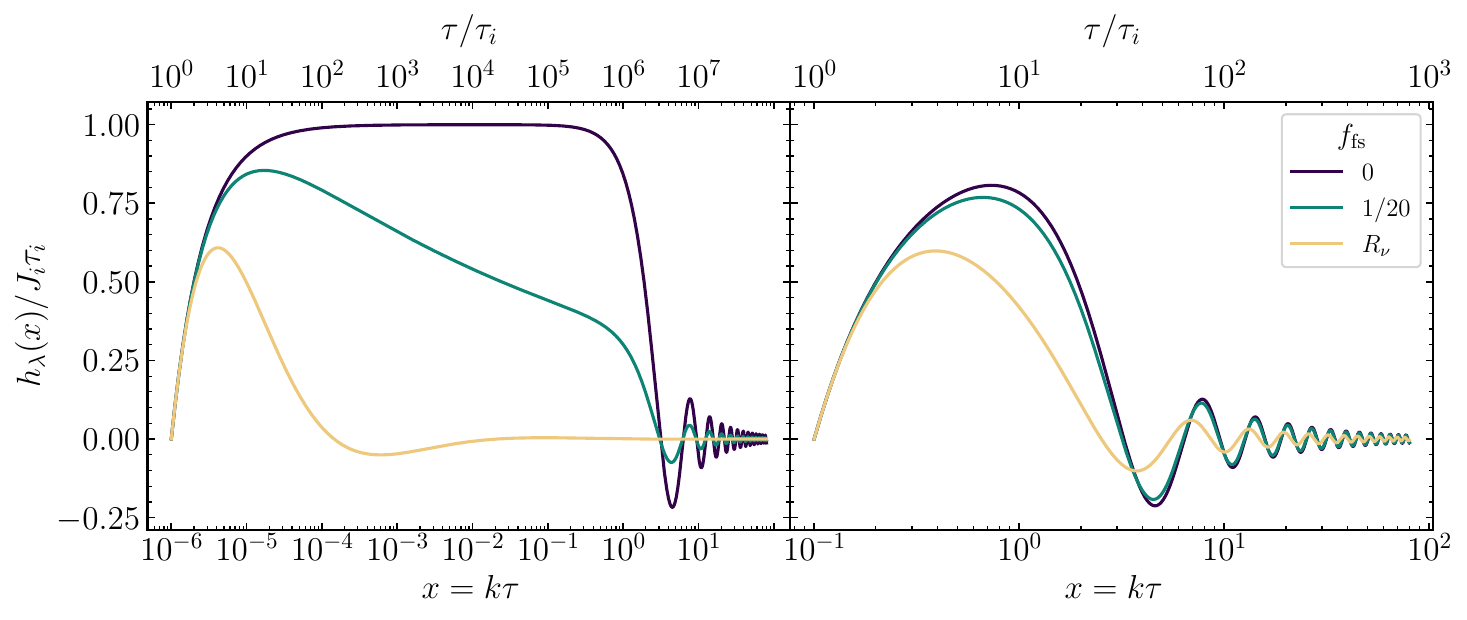}
    \caption{
        Dynamics of causal gravitational waves (in a radiation Universe) with wavelengths $10^6$
        (left panel) and $10$ (right panel) times larger than the horizon at the time of the phase
        transition, $\tau_i$.
        Each curve fixes a different abundance of free-streaming radiation $f_\mathrm{fs}$ as
        labeled in the legend.
        Note that the bottom axes are in units of $1/k$, i.e., the time of horizon crossing, while
        the top ones are in units of $\tau_i$, the time of the phase transition.
        Results are normalized by $J_i / \mathcal{H}(\tau_i) = J_i \tau_i$, the maximal amplitude
        that sufficiently superhorizon modes freeze out to.
    }
    \label{fig:causal-dynamics}
\end{figure}

\Cref{eqn:causal-sol} provides a decent approximation to full numerical solutions, the latter of
which we display in \cref{fig:causal-dynamics}.
By $\tau \sim 5 \tau_i$, free-streaming radiation reduces the gravitational wave amplitude first
attained after the kick---by as much as a factor of two for $f_\mathrm{fs} = 1$.
After this point, the dynamics are qualitatively dependent on the abundance of free-streaming
radiation.
When the Universe comprises entirely fluidlike radiation ($f_\mathrm{fs} = 0$), modes with
sufficiently small wavenumber $k$ freeze as discussed above, after which point they behave like
inflationary gravitational waves.
On the other hand, if $k \tau_i \gtrsim 10^{-1}$, as in the right-hand panel of
\cref{fig:causal-dynamics}, modes reenter the horizon before fully freezing, i.e., have nonzero
amplitude and velocity when they begin to oscillate.

In stark contrast, in the presence of free-streaming radiation gravitational waves evolve while
outside the horizon.
To understand the dynamics in \cref{fig:causal-dynamics}, observe that in the superhorizon limit of
the equation of motion [\cref{eqn:integral-equation-superhorizon}] free-streaming particles appear
to induce a mass for gravitational waves proportional to the Hubble scale.
Indeed, after jumping up to its maximal amplitude, the tensor perturbation slowly rolls in this
effective potential.
The superhorizon damping is therefore stronger on larger scales, as the tensor perturbation rolls
more toward zero the longer it remains outside the horizon (evident in comparing the two panels of
\cref{fig:causal-dynamics}).
For small free-streaming fractions $f_\mathrm{fs} \lesssim 10^{-1}$ (and so small effective masses),
the effect is monotonic, but for larger values the gravitational wave can in fact cross zero while
still outside the horizon.
For $f_\mathrm{fs} \gtrsim 1/2$, it completes one highly damped oscillation
while outside the horizon before decaying toward increasingly small values.
These results are summarized by the empirical results~\cite{Hook:2020phx, Brzeminski:2022haa}
\begin{align}\label{eqn:causal-damping-factor-vs-fs}
    \frac{\Omega_\mathrm{GW}}{\Omega_\mathrm{GW}[f_\mathrm{fs} = 0]}
    &\approx
        \begin{cases}
            \left( k \tau_i \right)^{16 f_\mathrm{fs} / 5}
            & f_\mathrm{fs} \lesssim 5/32
            \\
            k \tau_i
            \left(
                C_1 + C_2 \sin \left[ \sqrt{32 f_\mathrm{fs} / 5 - 1} \ln(k \tau_i) + C_3 \right]
            \right)
            & f_\mathrm{fs} \gtrsim 5/32
        \end{cases}
\end{align}
where the $C_a$ are mildly $f_\mathrm{fs}$-dependent factors.

\subsection{Results in interacting scenarios}\label{sec:results-causal-interacting}

We now turn to the effect of decoupling or recoupling self-interactions.
Since the superhorizon dynamics are well described by a simple ordinary differential equation
[\cref{eqn:integral-equation-superhorizon}], analytic results for instantaneous transitions could
feasibly be obtained  by matching solutions between the interacting and noninteracting regimes.
We instead simply discuss numerical results for self-interactions mediated by heavy ($n = 5$)
and light ($n = 1$) mediators as considered previously.

The suppression of causal gravitational waves depends on the precise timing of de-/recoupling
relative to the time of the phase transition, when the mode would reach its superhorizon, frozen
amplitude, and when it reenters the horizon.
The results are therefore complex and at times can be moderately sensitive to the precise timing
of de-/recoupling relative to one of the other relevant timescales.
We now summarize the relevant regimes and present results for a set of representative scenarios,
focusing on the effect of the timing of de-/recoupling relative to the time of the phase transition,
$\tau_\star / \tau_i$.
\begin{figure}[t!]
    \centering
    \includegraphics[width=\textwidth]{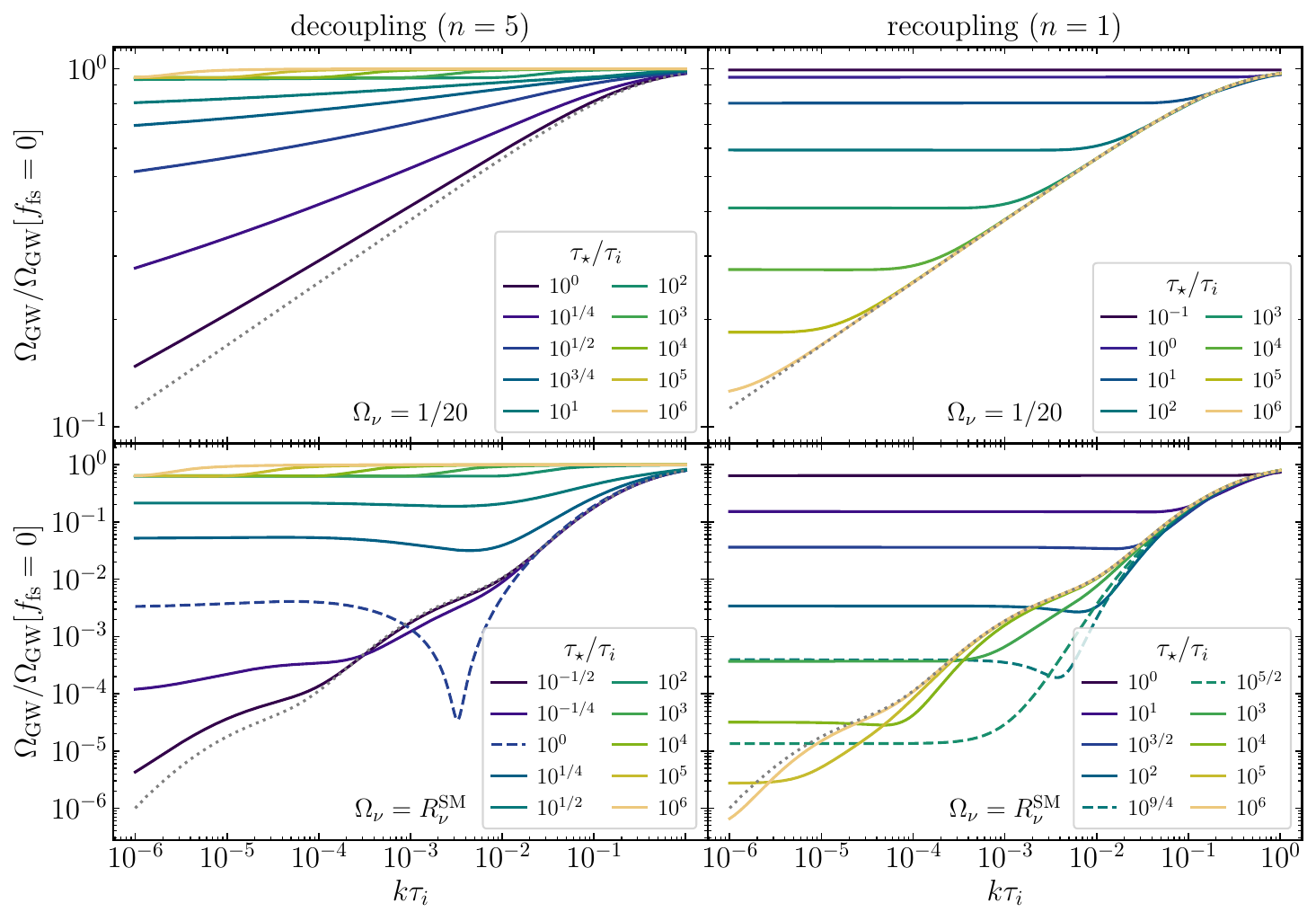}
    \caption{
        Damping factor for causal gravitational waves in scenarios with interacting radiation.
        The left and right panels respectively depict decoupling ($n=5$) and recoupling ($n=1$)
        scenarios.
        The top panels set the abundance of the relativistic species to $\fnu = 1/20$, while the
        bottom take $\fnu = \Rnu$, i.e., that of neutrinos after electron-positron annihilation
        [\cref{eqn:R-nu-standard-model}].
        Each panel plots results for a set of de-/recoupling times relative to the phase transition
        ($\tau_\star / \tau_i$) with colors labeled by each legend as well as that for the fully
        noninteracting case (dotted grey).
        Some curves in the bottom panels are dashed to aid in distinguishing their unique behavior.
        The damping factor is normalized to the result for a scenario without free-streaming
        particles and plotted as a function of wavenumber $k$ in units of the horizon size at
        the phase transition, $1/\tau_i$.
    }
    \label{fig:damping-causal-heavy-and-light-mediator}
\end{figure}
\Cref{fig:damping-causal-heavy-and-light-mediator} displays the squared damping factor for
all scenarios considered, each of which we discuss in turn.
Note that the quantity plotted in \cref{fig:damping-causal-heavy-and-light-mediator} multiplies
the spectrum in \cref{eqn:omega-gw-causal-general-eos}, i.e., acts as a scale-dependent
suppression on top of the standard $k^3$ dependence of $\Omega_\mathrm{GW}$ for modes that were
superhorizon during the phase transition (assuming a radiation-dominated Universe).

We first consider the decoupling of interactions, depicted in the left panels of
\cref{fig:damping-causal-heavy-and-light-mediator}.
We consider two concrete scenarios: one where the relativistic species contains $\fnu = 1/20$th of
the energy density in the Universe and another with $\fnu = 0.40523$, as would be relevant for,
e.g., dark phase transitions occurring around the time neutrinos decouple from the weak
interactions.
These would respectively belong to the small- and large-$f_\mathrm{fs}$ regimes of
\cref{eqn:causal-damping-factor-vs-fs}.
Naturally, if decoupling occurs before the phase transition ($\tau_\star < \tau_i$), the results
match the noninteracting scenario, as evident in \cref{fig:damping-causal-heavy-and-light-mediator}
where the darkest purple curves (with smallest $\tau_\star / \tau_i$) resemble the dotted grey
(depicting the fully noninteracting case).
The noninteracting result is not precisely reproduced for $\tau_\star$ slightly smaller than
$\tau_i$ merely because decoupling is not instantaneous (for the $T^5$ interaction rate employed).
For $\tau_\star / \tau_i \lesssim 10^{-1}$ or so, the imprint of the decoupling transition would be
nearly negligible (similar to the inflationary case where, per
\cref{fig:damping-vs-no-nu-vary-n-rad}, all modes with $k / k_\star \lesssim 10^{-1}$ are damped to
the same degree).

The next-simplest regime is that which behaves in exact analogy to inflationary gravitational waves:
if $\tau_\star \gtrsim 10^{3/2} \tau_i$, then all superhorizon modes reach the constant-amplitude
solution before decoupling.
The onset of free streaming at $\tau_\star$ has no immediate effect because the gravitational wave's
velocity is too small to source anisotropic stress, but indeed damps any modes that subsequently
enter the horizon (apparent in the green through yellow curves in
\cref{fig:damping-causal-heavy-and-light-mediator}).

When decoupling occurs sooner after the phase transition
($\tau_i < \tau_\star \lesssim 10^{3/2} \tau_i$), the results are more complex.
Efficient interactions suppress the neutrinos' response to the ``kick'' (i.e., the nonzero
gravitational-wave velocity), and, once they decouple, anisotropic stress can be sourced
and damp gravitational waves.
Because the gravitational wave is strongly damped after being kicked, its velocity is reduced at the
time of decoupling and thus sources less anisotropic stress.
For $\fnu = 1/20$, \cref{fig:damping-causal-heavy-and-light-mediator} makes apparent that
the net effect is a reduced tilt in the damping effect, mimicking the effect of a parametrically
smaller value of $f_\mathrm{fs}$ with no other scale-dependent features.
The results are substantially more complex for larger $\fnu$ (as in the bottom-left panel of
\cref{fig:damping-causal-heavy-and-light-mediator}) because of the oscillatory superhorizon
dynamics.
Intriguingly, if decoupling occurs near the time of the phase transition
($\tau_\star \approx \tau_i$), a prominent dip appears in the spectrum in the range
$10^{-3} \lesssim k \tau_i \lesssim 10^{-2}$.
The solutions are difficult to describe in this case, as modes in this range reenter the horizon
with varying amplitude and velocity.

We now turn to the recoupling scenarios depicted in the right panels of
\cref{fig:damping-causal-heavy-and-light-mediator}, considering the same values of $\fnu$.
For recoupling occurring just before the phase transition ($\tau_\star \lesssim \tau_i$), a small
and nearly scale-independent amplitude suppression occurs because interactions do not become
efficient instantaneously.
For later recoupling, the effective mass vanishes simultaneously for all modes that are still
superhorizon.
As a result, the damping factor is scale independent for wavenumbers outside the horizon at
recoupling, leading to a broken power law in the regime of small free-streaming fraction as
evident in the top-right panel of \cref{fig:damping-causal-heavy-and-light-mediator}.

In the large-$\fnu$ regime, the amplitude suppression still interpolates between that of
the noninteracting scenario (for $k \gg k_\star$) and a constant suppression (for $k \ll k_\star$),
but the intermediate regime can be more complex.
In particular, for $\tau_\star \gtrsim 10 \tau_i$ recoupling occurs while superhorizon modes are
oscillating due to the induced effective mass, at which point they freeze out.
The precise timing of these oscillations and recoupling again leads to a nontrivial spectral shape:
modes with $k \sim k_\star$ may end up more damped than if interactions had persisted (thereby
allowing them to continue to evolve outside the horizon).
For example, when $\tau_\star = 10^{9/4} \tau_i$ the damping factor exhibits a modest dip near
the horizon scale at recoupling.
For $\tau_\star = 10^{10/4} \tau_i$, recoupling occurs when all modes still outside the horizon are
at a point in their superhorizon oscillation of especially low amplitude.
After recoupling these modes freeze at that amplitude, incurring a net damping one to two orders of
magnitude greater than that if recoupling had occurred slightly earlier or later.
In sum, the interplay of recoupling and the unique superhorizon evolution of causal gravitational
waves coupled to free-streaming radiation permits conspicuous features in the (otherwise
near--power-law) spectrum at low frequencies, without requiring any particular tuning of the times
of recoupling and the phase transition.

\section{Discussion}\label{sec:discussion}

In this paper we investigated the gravitational-wave signatures of the interaction history of
relativistic particles.
If a nonnegligible fraction of the Universe's energy budget resides in radiation with
self-interactions (whether the Standard Model neutrinos or species from dark sectors), the
transition between strongly and weakly interacting regimes imprints characteristic features in
primordial gravitational wave backgrounds.
For gravitational waves seeded during inflation, the effect takes the form of a sigmoidal modulation
in amplitude, since the anisotropic stress generated by collisionless or free-streaming particles
damps gravitational waves as modes reenter the horizon.
On the other hand, if subhorizon dynamics (such as a first-order phase transition) source a
substantial gravitational wave background in the early Universe, superhorizon gravitational waves
(at frequencies well below the peak of the signal) are also generated and respond to the presence of
free-streaming particles even before horizon crossing.
Both of these scenarios permit model-agnostic descriptions of gravitational waves seeded outside
the cosmological horizon.
Because the coupling of gravitational waves to relativistic particles decays inside the horizon,
these two cases combine to provide a nearly comprehensive treatment of the imprint of the decoupling
or recoupling of interactions.\footnote{
    One source of near--horizon-sized gravitational waves not captured by the inflationary or
    causal descriptions is the anisotropic stress induced by scalar metric perturbations at second
    order in perturbation theory~\cite{Ananda:2006af,Baumann:2007zm,Domenech:2021ztg}.
    Refs.~\cite{Baumann:2007zm,Saga:2014jca} computed the signal on large scales sourced by the
    standard adiabatic perturbations, accounting for photon and neutrino anisotropic stress.
    The possibility of primordial black hole dark matter has motivated much interest in
    gravitational waves induced by enhanced density fluctuations on small scales (see
    Ref.~\cite{Domenech:2021ztg} and reference therein).
    The effect of anisotropic stress---which affects scalar perturbations at first order as
    well---would likely be difficult to account for in a model-independent manner, but was recently
    computed for a monochromatic spike in scalar power in Ref.~\cite{Zhang:2022dgx}.
}

\begin{figure}[t]
    \centering
    \includegraphics[width=\textwidth]{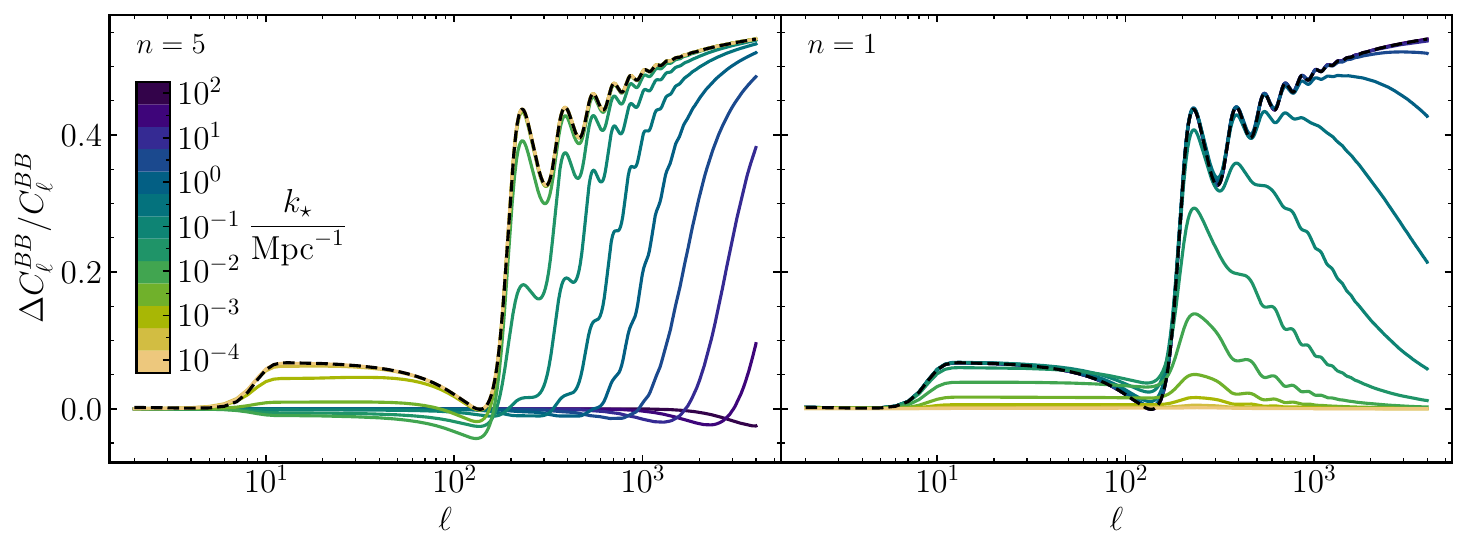}
    \caption{
        Relative change in the $B$-mode spectrum (compared to the SM case) in scenarios where
        nonstandard neutrino self-interactions decouple ($n = 5$, left) and recouple ($n = 1$,
        right).
        The colored curves vary the horizon size at de-/recoupling $k_\star$ from
        $10^{-4} \, \mathrm{Mpc}^{-1}$ to $10^{2} \, \mathrm{Mpc}^{-1}$ as indicated by the
        colorbar.
        The dashed black curves display the limiting case of a Universe without neutrinos (but with
        the radiation density unchanged).
    }
    \label{fig:cmb-b-modes}
\end{figure}

We now consider possible avenues for observing the signature of decoupling or recoupling interactions
in the gravitational wave background.
The primordial background from inflation, though well motivated, has yet eluded observation.
In the near term, the most promising means to detect inflationary gravitational waves is via the
$B$-mode polarization of the CMB, which would be sensitive to neutrino self-interactions that become
(in)efficient when observable modes enter the horizon.
In \cref{fig:cmb-b-modes} we display the relative change in the $B$-mode angular power spectrum for
representative decoupling and recoupling scenarios, computed by modifying
\textsf{CLASS}~\cite{Blas:2011rf} to include the interactions described in \cref{sec:interactions}.
Compared to the SM result, $B$-mode power is enhanced by as much as $\sim 50\%$ at small scales if
neutrinos are fluidlike at early times.
(The effect would be proportionally smaller if only one or two neutrino flavors interact, as in
\cref{fig:damping-heavy-light-partial}, or if considering radiation from a dark sector.)
The same transition in amplitude observed in the gravitational wave power spectrum
(\cref{fig:damping-heavy-light}) is evident over a finite range of multipoles, a steplike feature
that would measure the horizon size at the time of decoupling or recoupling.
These results generalize those of Ref.~\cite{Ghosh:2017jdy} (at a phenomenological level), which
considered two specific models of neutrino--dark-matter interactions.

In addition, if interactions are important at sufficiently late times, a phase offset in the tensor
modes' temporal evolution (due to the presence or absence of anisotropic stress) shifts the analog
of the acoustic peaks in the temperature and $E$-mode polarization spectra.
More generally, the locations of the peaks in the primordial $B$-mode spectrum in principle provide
a measure of the abundance of free-streaming vs. fluidlike radiation just as the acoustic peaks
do~\cite{Baumann:2015rya}.
Such features could help disentangle the effects of neutrino interactions from the amplitude and
tilt of the primordial tensor power spectrum.
Furthermore, given the current mild discord between \textit{Planck} and ACT in searches for evidence
of strongly interacting neutrinos~\cite{Kreisch:2022zxp,Corona:2021qxl} and early dark
energy~\cite{Hill:2021yec,Poulin:2021bjr,LaPosta:2021pgm,Smith:2022hwi}, information from tensor
perturbations (which depend on comparatively simpler physics than scalars) could provide a crucial
independent probe.

\begin{figure}[t]
    \centering
    \includegraphics[width=\textwidth]{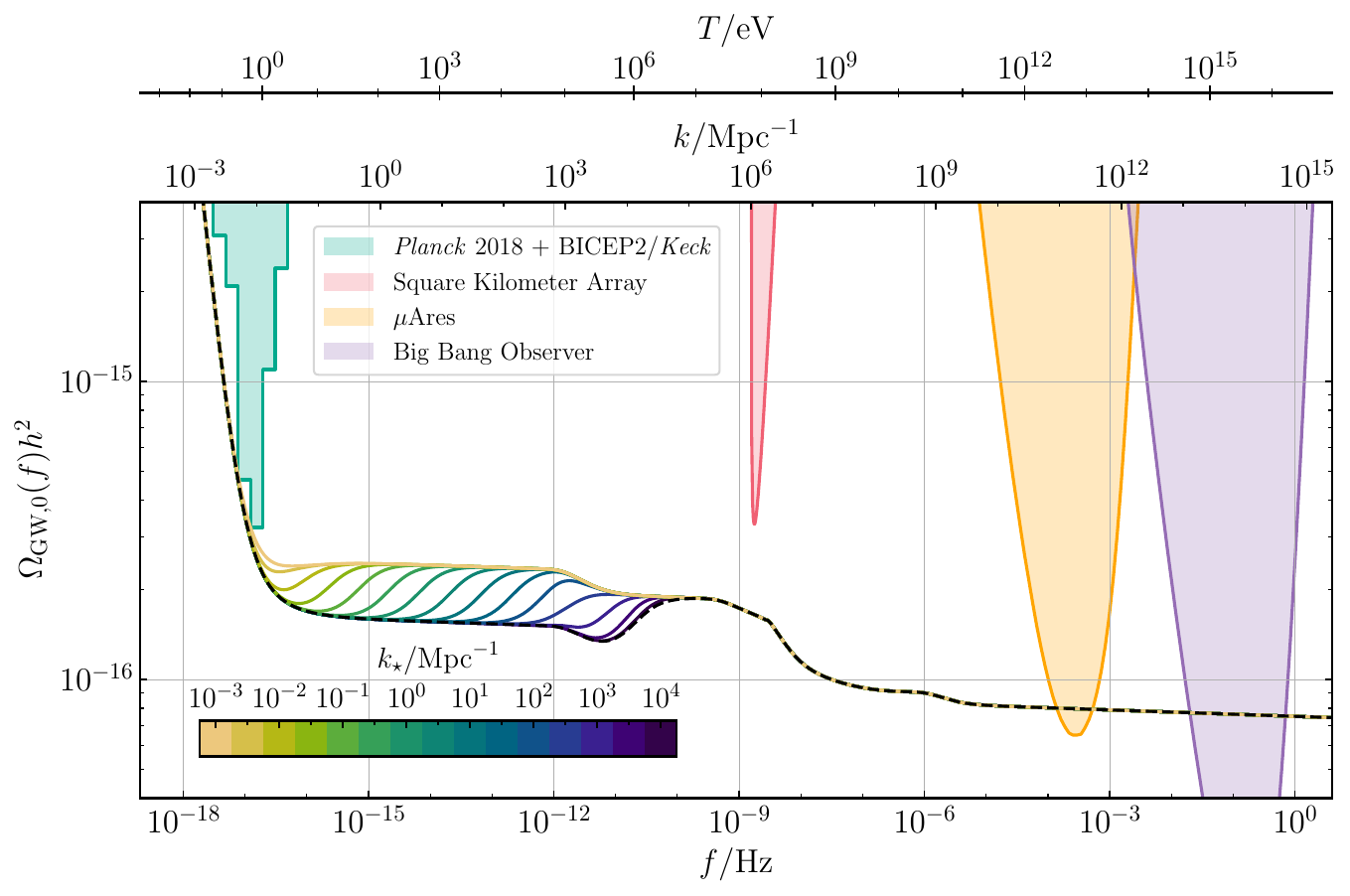}
    \caption{
        Spectral energy density in inflationary gravitational waves in decoupling scenarios
        that delay neutrino free streaming relative to the decoupling of the weak interactions.
        Colored curves display results for scenarios with interaction rate proportional to $T^5$
        and various strengths parameterized in terms of the horizon size at decoupling (indicated by
        the colorbar), while the black dashed curve corresponds to the Standard Model result.
        All results assume single-field slow roll inflation with $r = 0.056$ and account for thermal
        history effects in \cref{eqn:gw-amplitude-transfer-function} using data from
        Ref.~\cite{Saikawa:2018rcs}.
        Superimposed are recent constraints from \textit{Planck} and
        BICEP2/\textit{Keck}~\cite{Planck:2018jri,BICEP2:2018kqh} as computed by
        Ref.~\cite{Clarke:2020bil}, forecasted power-law--integrated sensitivities of the Square
        Kilometre Array~\cite{Carilli:2004nx,Janssen:2014dka,Weltman:2018zrl} and the proposed
        Big-Bang Observer~\cite{Crowder:2005nr,Corbin:2005ny,Harry:2006fi}, both, provided by
        Refs.~\cite{Schmitz:2020syl,schmitz_kai_2020_3689582}, and that of the proposed $\mu$Ares
        detector~\cite{Sesana:2019vho}, with color labeled by the legend.
    }
    \label{fig:omega-gw-decoupling}
\end{figure}

Further in the future, pulsar timing and interferometers will provide a high-frequency lever arm for
measuring an inflationary gravitational wave background, jointly limiting the amplitude and tilt of
the inflationary spectrum~\cite{Lasky:2015lej}.
Measurements at disparate scales would also probe the abundance of free-streaming radiation over a
large span of temperatures~\cite{Chacko:2015noa, Liu:2015psa}, a possibility we consider in
\cref{fig:omega-gw-decoupling}.
Though the dearth of proposed methods for observing gravitational waves near picohertz frequencies
is unfortunate for the prospects of probing neutrino decoupling,\footnote{
    CMB spectral distortions are sensitive to gravitational waves between femtohertz and nanohertz
    frequencies~\cite{Chluba:2014qia, Kite:2020uix}; futuristic experiments could provide a
    complementary probe of, e.g., the peak of causal signals.
} observations at higher frequencies---whether of inflationary or causal gravitational waves---would
probe new particles that decouple from the SM plasma while relativistic.
Current constraints limit such light relics to comprise $\lesssim 4\%$ of the radiation budget at
the time of the CMB, but their relative contribution would be roughly twice as large at temperatures
above the electroweak scale~\cite{Dvorkin:2022jyg}.
Larger abundances of free-streaming radiation in the very early Universe can be accommodated by more
exotic cosmologies (if, say, that radiation later decays, or other degrees of freedom inject entropy
into the SM plasma).

In more speculative scenarios, the low-frequency gravitational wave signal from causal, subhorizon
processes, such as phase transitions or resonant particle production, would be sensitive to the
interaction history of radiation.
The low-frequency tail (i.e., modes that were superhorizon at the time of production) is tilted blue
(relative to the typical $f^3$ dependence) under the influence of tensor anisotropic stress;
the decoupling or recoupling of interactions in relativistic species would lead to a break in the
power law, evident in \cref{fig:damping-causal-heavy-and-light-mediator}.
If the abundance of free-streaming radiation is large ($f_\mathrm{fs} \gtrsim 5/32$), additional
oscillatory features arise.
Ref.~\cite{Brzeminski:2022haa} demonstrated that future space-borne interferometers could
optimistically probe, through measurements of the causal tail from phase transitions, the thermal
history of the Universe at the percent or permille level---including the effect of a single new
light, free-streaming particle.
The effects of decoupling or recoupling observed in
\cref{fig:damping-causal-heavy-and-light-mediator} could likely be probed to the same degree.
Finally, dark sector dynamics that source gravitational waves after neutrino decoupling but before
recombination (as motivated by, e.g., early dark
energy~\cite{Niedermann:2020dwg,Weiner:2020sxn,Niedermann:2021vgd} and the
axiverse~\cite{Kitajima:2018zco,Cyncynates:2022wlq}) could produce signals visible to future CMB
experiments that would be sensitive to whether neutrinos interact or free-stream.
The discovery of gravitational wave backgrounds from such novel scenarios (or from inflation or
phase transitions) would provide an exciting opportunity to learn about not just the physics
underlying their sources but also possible nonstandard neutrino interactions, dark radiation
sectors, and the dynamics of the early Universe over a broad range of energy scales.

\acknowledgments

This research is supported by the Department of Physics and the College of Arts and Sciences at the
University of Washington.
ML is grateful for support from the Dr. Ann Nelson Endowed Professorship in Physics.
This work made use of the Python packages \textsf{NumPy}~\cite{Harris:2020xlr},
\textsf{SciPy}~\cite{Virtanen:2019joe}, \textsf{matplotlib}~\cite{Hunter:2007ouj},
\textsf{SymPy}~\cite{Meurer:2017yhf}, \textsf{mpmath}~\cite{mpmath}, and
\textsf{CMasher}~\cite{cmasher}.

\appendix

\section{Numerical implementation}\label{app:numerical-implementation}

In this appendix we outline the numerical implementation of the various calculations performed in
this work.
Regardless of the background evolution (for radiation and matter-radiation Universes) or of the
initial conditions (inflationary or causal), we solve \cref{eqn:eoms-ito-x} with the Boltzmann
hierarchy for $F^{(T)}$ truncated at $l_\mathrm{max} = 100$ with the truncation prescription of
Ref.~\cite{Ma:1995ey},
\begin{align}
    F_{\lambda, l_\mathrm{max}}
    \approx \frac{2 l_\mathrm{max} - 1}{k \tau} F_{\lambda, l_\mathrm{max} - 1}
        - F_{\lambda, l_\mathrm{max} - 2}.
\end{align}
We use \textsf{SciPy}'s~\cite{Virtanen:2019joe} \texttt{solve\_ivp} routine to solve the system
numerically.
For noninteracting scenarios, we use the explicit Runge-Kutta method
DOP853~\cite{Hairer1993nonstiff} with relative tolerance $10^{-6}$ and absolute tolerance
$10^{-12}$.
The self-interaction terms, on the other hand, make the system stiff and better suited to implicit
methods; in this case we use the Radau IIA routine~\cite{Hairer1993stiff} with the Jacobian of
\cref{eqn:eoms-ito-x} computed by hand (i.e., analytically) as a sparse matrix.
To reduce the sensitivity of the adaptive time stepping to the exponentially damped $F_{\lambda, l}$
(whenever interactions are efficient), we increase its absolute tolerance to $10^{-9}$ and decrease
that of $\hD_\lambda$ and $\partial_x \hD_\lambda$ to $10^{-15}$.

In all cases, we evolve the system to $x_f \equiv k \tau_f = 100$; the chosen $l_\mathrm{max} = 100$
safely ensures that errors from the truncation of the Boltzmann hierarchy do not propagate back to
the first few moments until $k \tau \sim 100$ (at which point the anisotropic stress itself is small
anyway).
To compute the damping coefficient $A_\infty$, we match on to the solution to \cref{eqn:h-eom-ito-x}
in a Universe with constant equation of state and no anisotropic stress,
\begin{align}
\begin{split}
    \hD_\lambda(x)
    &= \frac{x_f^{\alpha + 1}}{x^{\alpha - 1}}
        \left(
            - j_{\alpha-1}(x)
            \left[
                y_{\alpha}(x_f)
                \hD_\lambda(x_f)
                + y_{\alpha-1}(x_f) \hD_\lambda'(x_f)
            \right]
        \right.
    \\
    &\hphantom{{}={} \frac{x_f^{\alpha + 1}}{x^{\alpha - 1}} \Bigg( }
        \left.
            + y_{\alpha-1}(x)
            \left[
                j_{\alpha}(x_f)
                \hD_\lambda(x_f)
                + j_{\alpha-1}(x_f) \hD_\lambda'(x_f)
            \right]
        \right).
\end{split}
\end{align}
Here primes denote derivatives with respect to $x$.
When both $x$ and $x_f$ are large,
\begin{align}\label{eqn:A-inf-x_f}
    A_\infty
    &\approx a(x_f / k) \sqrt{
        \hD_\lambda(x_f)^2
        + \hD_\lambda'(x_f)^2
    }.
\end{align}
In a single component Universe, $a(x_f/k) = x_f^\alpha$ (normalized to the scale factor at horizon
crossing) and \cref{eqn:A-inf-x_f} holds asymptotically (and at all times in radiation Universes).
These results are also accurate in matter-radiation Universes if evaluated when modes are deep
inside the horizon, since the Hubble rate then changes slowly compared to the gravitational-wave
oscillation frequency.
We have verified that these approximations achieve subpercent accuracy in evaluating the relative
effect of neutrinos compared to Universes with only fluidlike radiation.
See, e.g., Refs.~\cite{Pritchard:2004qp,Watanabe:2006qe,Weinberg:2008zzc} for more formal procedures
to match analytic solutions between the radiation and matter eras.

To compute the $B$-mode power spectrum of the CMB, we modify \textsf{CLASS} to include neutrino
interactions in the same manner.
(\textsf{CLASS} of course implements a complete $\Lambda$CDM cosmology that accounts for dark energy
at late times.)
\textsf{CLASS}'s implicit solver is less robust to the equations becoming extremely stiff (e.g., at
very early times for decoupling scenarios), so we cap the maximum value of the interaction rate to
$10^{12} \, \mathrm{Mpc}^{-1}$ and verified that increasing or decreasing this value had a
negligible impact on the results.

\section{Semianalytic results for inflationary gravitational waves}\label{app:semianalytic-inflation}

This appendix outlines the computation that yields the semianalytic results for the amplitude and
phase of inflationary gravitational waves in the presence of free-streaming particles,
\cref{eqn:amplitude-sq-analytic,eqn:phase-analytic}.

Following Ref.~\cite{Dicus:2005rh, Shchedrin:2012sp, Stefanek:2012hj}, express the convolution of
spherical Bessel functions as a series thereof:
\begin{align}\label{eqn:def-alpha-n}
    - \int_0^x \ud u \, K(x - u) \partial_u j_0(u)
    &= \sum_{n = 0}^\infty c_n j_n(x).
\end{align}
[Recall that $\partial_u j_0(u) = - j_1(u)$.]
Eq.~(24) of Ref.~\cite{Shchedrin:2012sp} provides an analytic result for the convolution in
\cref{eqn:def-alpha-n} in terms of Clebsch-Gordan coefficients, which we evaluate using
\textsf{SymPy}~\cite{Meurer:2017yhf}.
Only the even $c_n$ are nonzero, due to symmetry properties of the Clebsch-Gordan coefficients
(see Refs.~\cite{Dicus:2005rh, Shchedrin:2012sp, Stefanek:2012hj} for details).

In order to compute the amplitude and phase [\cref{eqn:amplitude-sq-analytic,eqn:phase-analytic}],
we must obtain $\mathcal{I}_s$ and $\mathcal{I}_c$ from \cref{eqn:I-cs-defn}, which in turn requires
integrating the order-zero spherical Bessel functions against spherical Bessel functions.
In terms of the hypergeometric function ${}_2{F}_3$, these evaluate to
\begin{subequations}
\begin{align}
    \int_0^x \ud \tilde{x} \,
        j_n(\tilde{x})
        y_0(\tilde{x})
    &= - \frac{\sqrt{\pi} x^{n}}{2^{n + 1} n \Gamma(n + 3/2)}
        {}_2{F}_3\left(
            \frac{n + 1}{2}, \frac{n}{2};
            \frac{1}{2}, n + 1, n + \frac{3}{2}; -x^2
        \right) \\
    \int_0^x \ud \tilde{x} \,
        j_n(\tilde{x})
        j_0(\tilde{x})
    &= \frac{\pi x^{n+1}}{2^{n + 2} \Gamma(3/2) \Gamma(n + 3/2) \Gamma(n + 2)}
        {}_2{F}_3\left(
            \frac{n + 1}{2}, \frac{n}{2} + 1;
            \frac{3}{2}, \frac{3}{2} + n, n + 2; -x^2
        \right).
\end{align}
\end{subequations}
For even $n > 0$, the series expansions of the above about large $x$ are
\begin{subequations}\label{eqn:analytic-integral-bessels}
\begin{align}
    \int_0^x \ud \tilde{x} \,
        j_n(\tilde{x})
        y_0(\tilde{x})
    &= - \frac{(-1)^{n/2}}{n (n + 1)}
        + \frac{1}{x^2} \frac{(-1)^{n/2}}{8} \left[ n (n + 1) + 2 \cos(2 x) \right]
        + \mathcal{O}(x^{-3}) \\
    \int_0^x \ud \tilde{x} \,
        j_n(\tilde{x})
        j_0(\tilde{x})
    &= - \frac{(-1)^{n/2}}{2 x}
        \left[
            1 +
            \frac{\cos x \sin x}{x}
        \right]
        + \mathcal{O}(x^{-3}).
\end{align}
\end{subequations}
We observe that $\abs{c_n} \sim n^{-5}$ for large $n$, so the series appearing in $\mathcal{I}_{s}$
and $\mathcal{I}_{c}$ [i.e., after substituting \cref{eqn:analytic-integral-bessels} into
\cref{eqn:I-cs-defn}] converge.
Numerically evaluating the required series (up to $256 - 512$ terms) yields
\begin{subequations}
\begin{align}
    24 \sum_{m = 1}^\infty
        (-1)^{m}
        c_{2 m}
    &\approx - 2 \\
    24 \sum_{m = 1}^\infty
        \frac{(-1)^{m}}{2 m (2 m + 1)}
        c_{2 m}
    &\approx - \frac{5}{9} \\
    24 \sum_{m = 1}^\infty
        (-1)^{m} \cdot 2 m (2 m + 1)
        c_{2 m}
    &\approx 4.
\end{align}
\end{subequations}
Therefore, to leading order in large $x$, \cref{eqn:I-cs-defn} evaluates to
\begin{subequations}
\begin{align}
    \mathcal{I}_s(x)
    &\approx - \hD_{\lambda, 0}
        f_\mathrm{fs}
        \left(
            \frac{5}{9}
            + \frac{1}{x^2} \sin^2 x
        \right) \\
    \mathcal{I}_c(x)
    &\approx
        - \hD_{\lambda, 0}
        \frac{f_\mathrm{fs}}{x}
        \left(
            1 +
            \frac{\cos x \sin x}{x}
        \right).
\end{align}
\end{subequations}
Substituting these asymptotic expansions into \cref{eqn:amplitude-ito-hs-hc,eqn:phase-ito-hs-hc}
yields \cref{eqn:amplitude-sq-analytic,eqn:phase-analytic}.

\bibliography{references}
\bibliographystyle{JHEP}

\end{document}